\documentclass[reprint,pre,aps, 
	footinbib 
	]{revtex4-2}
    	
\usepackage[english]{babel}
\usepackage[T1]{fontenc}
\usepackage[utf8]{inputenc}
\usepackage{lmodern}
\usepackage{amssymb}
\usepackage{mathtools}
\usepackage{bbm}
\usepackage{verbatim}
\usepackage[dvipsnames]{xcolor}
\usepackage{graphicx}
\usepackage{tikz}
    \usetikzlibrary{decorations.pathreplacing}
    \usetikzlibrary{decorations.pathmorphing} 
    \usetikzlibrary{matrix,arrows}
\usepackage{multirow}
\usepackage{enumitem}
    	
\usepackage[ocgcolorlinks, linkcolor=BlueViolet, citecolor=OliveGreen, urlcolor=RawSienna, filecolor=Sepia, hyperfootnotes=false]{hyperref}

\usepackage{xcolor}
\usepackage{bm}

\newcommand{\dd}{\mathrm{d}}
\newcommand{\CO}{{\cal O}}
\newcommand{\be}{\begin{align}}
\newcommand{\ee}{\end{align}}

\newcommand{\ii}{\mathrm{i}}

\newcommand{\p}{\partial}
\newcommand{\K}{\mathrm{K}}
\newcommand{\STr}{\operatorname{STr}}
\newcommand{\tE}{\widetilde{E}}

\newcommand{\tmu}{{\mu}}
\newcommand{\tH}{H}
\newcommand{\tK}{k}
\newcommand{\Sm}{\mathcal{S}}
\newcommand{\Pf}{\mathrm{Pf}}
\newcommand{\mf}{\mathfrak{f}}

\def\no{\nonumber}
\def\bbB{{\mathbb B}}
\def\H{{\rm H}}
\def\CG{{\cal G}}
\def\CS{{\cal S}}
\def\Tr{{\rm Tr}}
\def\u{{\bf u}}
\def\v{{\bf v}}
\def\w{{\bf w}}
\def\twist{\gamma}

\begin{document}

\title{Exact Three-Point Functions in $\mathcal{N}=2$ Superconformal Field Theories:\\ Integrability vs. Localization}
    	
\author{Gwena\"el Ferrando\textsuperscript{$a$}, Shota Komatsu\textsuperscript{$b$}, Gabriel Lefundes\textsuperscript{$c \, d$} and Didina Serban\textsuperscript{$c$} }
\affiliation{\textsuperscript{$a$}\,Bethe Center for Theoretical Physics, Universität Bonn,
    Wegelerstr. 10, D-53115, Germany}
\affiliation{\textsuperscript{$b$}\,CERN Department of Theoretical Physics, CERN
    1211 Meyrin, Switzerland}
\affiliation{\textsuperscript{$c$}\,Universit\'e Paris--Saclay, CNRS, CEA, Institut de Physique Th\'eorique, 91191 Gif-sur-Yvette, France} 
\affiliation{\textsuperscript{$d$}\,ICTP South American Institute for Fundamental Research, IFT-UNESP, S\~ao Paulo, SP 01440-070, Brazil}
    	
\begin{abstract}
\noindent 
     We propose an integrability approach for planar three-point functions at finite coupling in  $\mathcal{N}=2$ superconformal field theories obtained as $\mathbb{Z}_K$ orbifolds of $\mathcal{N}=4$ super Yang-Mills (SYM). Generalizing the hexagon formalism for $\mathcal{N}=4$ SYM, we reproduce the structure constants of Coulomb branch operators, previously obtained by supersymmetric localization, as exact functions of the 't Hooft coupling. Our analysis explains the common physical origin of Fredholm kernels in integrability and localization, and hints at structures after the resummation in the hexagon formalism.
\end{abstract}
\hfill BONN-TH-2025-09
    
\maketitle
    
\noindent{\textbf{Introduction.}} 
The understanding of strongly-coupled gauge theory has advanced significantly thanks to dualities, holography and non-perturbative methods like integrability, localization and bootstrap. The paradigmatic theory in which this progress has been made is ${\cal N}=4$ supersymmetric Yang--Mills (SYM) in four dimensions; the most symmetric and thus the most tractable. 
    
Studying theories with less symmetries has proven useful in testing the applicability of these methods and uncovering more general structures. Notably, there exists a broad class of $\mathcal{N}=2$ supersymmetric theories with rich physical and mathematical structures (see e.g.~\cite{Seiberg:1994aj,Seiberg:1994rs,Gaiotto:2009we}). Among them, $\mathbb{Z}_K$-orbifiolds of $\mathcal{N}=4$ SYM \cite{Kachru:1998ys,Gukov:1998kk} are particularly interesting, as they possess conformal invariance and integrability \cite{Beisert:2005he,Solovyov:2007pw,Gadde:2010zi} despite being less supersymmetric. The planar spectrum of this theory was studied extensively by integrability \cite{Gadde:2010ku,Arutyunov:2010gu,Beccaria:2011qd,deLeeuw:2012hp,Skrzypek:2022cgg}, and more recently, correlation functions of BPS operators (known as Coulomb-branch operators) were computed exactly using supersymmetric localization \cite{Beccaria:2021hvt, Beccaria:2022ypy,  Billo:2022xas, Billo:2022lrv, Billo:2022fnb, Billo:2022gmq, Bajnok:2024epf, Bajnok:2024ymr, Bajnok:2024bqr, Korchemsky:2025eyc}. In the planar limit, these results, involving operators in the so-called twisted sector, are given by Fredholm determinants of integrable Bessel operators. Surprisingly, similar expressions arise in the integrability approach to various  observables of $\mathcal{N}=4$ SYM \cite{Fleury:2016ykk,basso2019,Basso:2020xts,Sever:2020jjx,Sever:2021nsq,Sever:2021xga,Basso:2023bwv}, such as large-charge four-point functions \cite{Coronado:2018ypq,Coronado:2018cxj,Bargheer:2019exp,Kostov:2019stn,Kostov:2019auq,Belitsky:2019fan,Belitsky:2020qrm,Belitsky:2020qir,Belitsky:2020qzm,Kostov:2021omc}. There, the Fredholm determinant arises as the partition function of magnon excitations on the worldsheet. This raises the natural questions of whether the localization results for $\mathcal{N}=2$ orbifolds can be reproduced by integrability and whether they, too, can be interpreted in terms of a partition function.
    
In this Letter, we give an affirmative answer to these questions by generalizing the hexagon formalism for three-point functions of $\mathcal{N}=4$ SYM  \cite{Basso:2015zoa} to account for orbifolding. A key challenge in this approach is handling the divergences from virtual magnons wrapping around each operator, which require systematic regularization  \cite{Basso:2017muf,Basso:2022nny}. We propose to perform the regularization using a genus-two surface (cf.~\cite{Basso_IGST22}). The procedure leads to an all-loop, all-magnon expression for the wrapping contribution in terms of a vertex-like partition function. Evaluating this expression up to three virtual  magnons we reproduce localization results. The result for an arbitrary number of magnons is left as a conjecture. More broadly, $\mathcal{N}=2$ orbifolds are an ideal setup for developing computational techniques of the hexagon formalism. Our formalism can be applied to non-BPS observables in these theories, and our analysis lays 
the basis for future studies \footnote{After the first version of this work was submitted on arXiv, the hexagon formalism was set up to work for non-BPS states at tree-level in \cite{lePlat:2025eod} }.

\medskip
    
\noindent{\textbf{Three-point functions from localization.}} We consider the $\mathcal{N}=2$ quiver gauge theory at the $\mathbb{Z}_K$ orbifold point, where the $K$ gauge couplings coincide. A convenient way to describe the theory is to start with SU$(K N)$ $\mathcal{N}=4$ SYM and perform an orbifold projection by $\twist={\textrm{diag}}({\bf{1}}_N, \rho \  {\bf{1}}_N\dots ,\rho^{K-1} \ {\bf{1}}_N)$ with $\rho \equiv e^{{2\pi \ii}/{K}}$,
\begin{align}
    \twist\,(A_\mu,Z)\,\twist^{-1} =  (A_\mu,Z)\, , \ \twist\, (X, Y )\,\twist^{-1} = \rho \,(X, Y)\, ,
\end{align}
where $X,Y$ and $Z$ are complex scalars and $A_\mu$ the gauge potential. After the projection, single-trace operators consist of an untwisted sector, taking the same form as in $\mathcal{N}=4$ SYM, and $K-1$ twisted sectors, given by insertions of powers of $\gamma$ in the trace. For instance, the untwisted and twisted BPS operators read
\begin{align}
\label{extop}
    \mathcal{O}^{(0)}_\ell =\frac{1}{\sqrt{K}} \Tr\, Z^\ell (x)\,, \ \,  \mathcal{O}^{(\alpha)}_\ell(x)=\frac{1}{\sqrt{K}} \Tr\, \gamma^\alpha Z^\ell (x)\, .
\end{align}
In the spin-chain language \cite{Beisert:2004ry,Beisert:2005he,Gadde:2010ku, Zoubos:2010kh}, the orbifold action can be realized by the insertion of a twist in the spin chain, which breaks the symmetry from $\textrm{PSU}(2,2|4)$ to $\textrm{PSU}(2,2|2) \times \textrm{SU}(2)$ for $K=2$ and to $\textrm{PSU}(2,2|2) \times \textrm{U}(1)$ for $K>2$. On the flavor indices of magnons over the $Z$ vacuum the twists are implemented by integer powers of ${\bf{1}}_L\times \tau_R={\bf{1}}_L\times(\rho,\rho^{-1},1,1)_R$, see the Supplemental Material for more details. The twist breaks the $\textrm{PSU}(2|2)^2$ symmetry of the $\mathcal{N}=4$ SYM magnons down to $\textrm{PSU}(2|2)\times \textrm{SU}(2)\times \textrm{U}(1)$.
    
The two- and three-point functions of BPS operators  were computed by localization in \cite{Beccaria:2021hvt} and checked perturbatively in \cite{Galvagno:2020cgq} and by supergravity in \cite{Billo:2022gmq,Skrzypek:2023fkr}. The results for normalized (extremal) three-point functions are
\begin{align}
\label{3ptBPS}
\begin{aligned}
    \frac{\langle \mathcal{O}^{(\alpha_1)}_k(x)\,  \mathcal{O}^{(\alpha_2)}_\ell(y)\,\bar {\mathcal{O}}^{(\alpha_3)}_p(z)\rangle}{\sqrt{\langle \mathcal{O}_k\bar{\mathcal{O}}_k\rangle\langle \mathcal{O}_{\ell}\bar{\mathcal{O}}_{\ell}\rangle\langle \mathcal{O}_p\bar{\mathcal{O}}_p\rangle}} &=\frac{\sqrt{k\ell p}}{\sqrt{K}N}\frac{C^{(\alpha_1,\alpha_2,\alpha_3)}_{k,\ell,p}}{|x-z|^{2k}|y-z|^{2\ell}} \,,
\end{aligned}
\end{align}
where $p=k+\ell$, $\alpha_3 = \alpha_1 + \alpha_2$ and  $g=\sqrt{g_{\rm YM}^2 N}/4\pi$. The structure constants $C^{(\alpha_1,\alpha_2,\alpha_3)}_{k,\ell,p}(g)$ take a factorized form
\begin{align}
\label{Cfact}
    &C^{(\alpha_1,\alpha_2,\alpha_3)}_{k,\ell,p} = C^{(\alpha_1)}_k \, C^{(\alpha_2)}_\ell \, C^{(\alpha_3)}_p \, , \\
    &C^{(\alpha)}_{L}=\sqrt{1+\frac{g}{2\ell}\p_g\ln\left(\frac{\det (1-s_{\alpha}K_{L+1})}{\det (1-s_{\alpha}K_{L-1})}\right)}\,,
\end{align}
where $K_{L-1}$ is a semi-infinite matrix with elements
\begin{align}
    \frac{(K_{L-1})_{mn}}{\sqrt{(2m\!+\!L)(2n\!+\!L)}}&\!=\!-8\int_0^\infty\! \frac{\dd t}{t}\chi_g(t)\,J_{2m+L}(t)\,J_{2n+L}(t),\no \\
    \chi_g(t)&={e^{t/2g}}/{(e^{t/2g}-1)^2}\,,
    \label{lockernel}
\end{align}
and $s_{\alpha} = \sin^2{{\pi \alpha}/{K}}$ is the character of the twist in the fundamental representation of PSU$(2|2)$. The kernel described above coincides with the octagon kernel for the large-charge four-point functions in \cite{Kostov:2019stn,Kostov:2019auq,Belitsky:2020qrm,Belitsky:2020qir}, with cross ratios set to particular values $\theta =\pi,\, \xi=\phi=\varphi=0$, or  $z=\bar z=1$ and $\alpha=\bar \alpha=-1$, in the notations of \cite{Kostov:2019auq,Belitsky:2020qir}. At weak coupling, they can be expanded as
\begin{align}
\label{Cdeterminantlimits}
    C^{(\alpha)}_L(g)=1+\CO(g^{2L})\,, \quad C^{(\alpha)}_L(0)=C^{(\alpha)}_\infty(g)=1\,.
\end{align}

We observe that the result obtained from localization can be rewritten in the following more transparent form \footnote{After we presented this form of the normalized structure constant \eqref{Cdeterminant} in \cite{Serban_IGST24}, G. Korchemsky informed us he obtained it independently, and it was published recently in \cite{Korchemsky:2025eyc}. This was then used in \cite{Korchemsky:2025mla} to derive a combinatorial expansion of $\det (1-s_{\alpha}K_{L})$ in terms of iterated Chen integrals.}:
\begin{equation}\label{Cdeterminant}
    C^{(\alpha)}_{L}=\frac{\det (1-s_\alpha K_{L})}{\sqrt{\det (1-s_{\alpha}K_{L-1})\,\det (1-s_{\alpha}K_{L+1})}}\, .
\end{equation}
Our main result, explained below, is to reproduce the expression \eqref{Cdeterminant} from integrability.
    
\medskip
    
\begin{figure}[t]
\centering
    \includegraphics[scale=0.46]{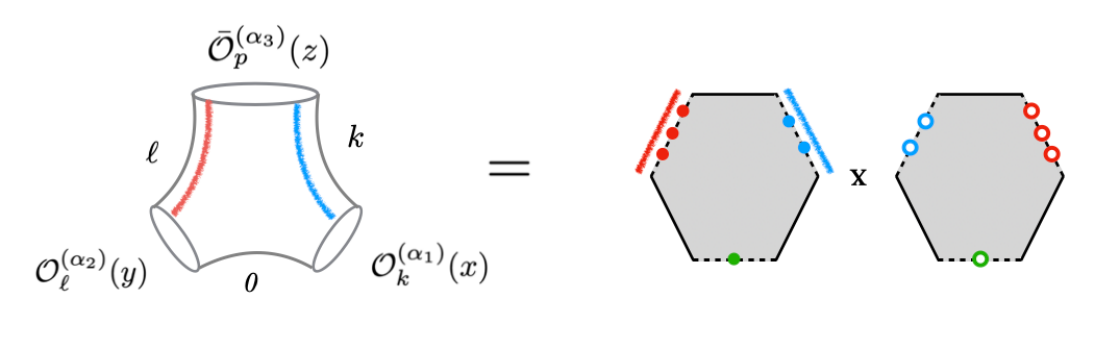}
    \caption{Structure constant corresponding to \eqref{3ptBPS}. The numbers on the sides of the pants denote the bridge lengths. The thick colored lines denote the twist insertions.}\label{fig:WrappedOctagon}
\end{figure}

\noindent{\textbf{Hexagon formalism for orbifold $\mathcal{N}=2$ SCFT.}}
In the integrability framework, the three-point function is represented pictorially as a pair of pants, which is then cut into two hexagonal tiles. The hexagon form factors were determined exactly at finite coupling through integrability \cite{Basso:2015zoa,Fleury:2016ykk,basso2019}. To glue the hexagons back together, one inserts complete sets of states on edges of the hexagons (called \textit{bridges}), as in Figure \ref{fig:WrappedOctagon}. The associated excitations, referred to as (virtual, or mirror) \textit{magnons}, propagate from one hexagon to the other with an exponential suppression factor that depends on their energy and the length of the  bridge.

We propose that this procedure can be extended to the $\mathbb{Z}_K$ orbifold theory by inserting powers of twists $\tau$ on bridges as illustrated in Figure \ref{fig:WrappedOctagon}. The distribution of twists on the bridges is dictated by the extremality of the three point function. 
    
\begin{figure*}[t]
\centering
    \includegraphics[scale=0.2]{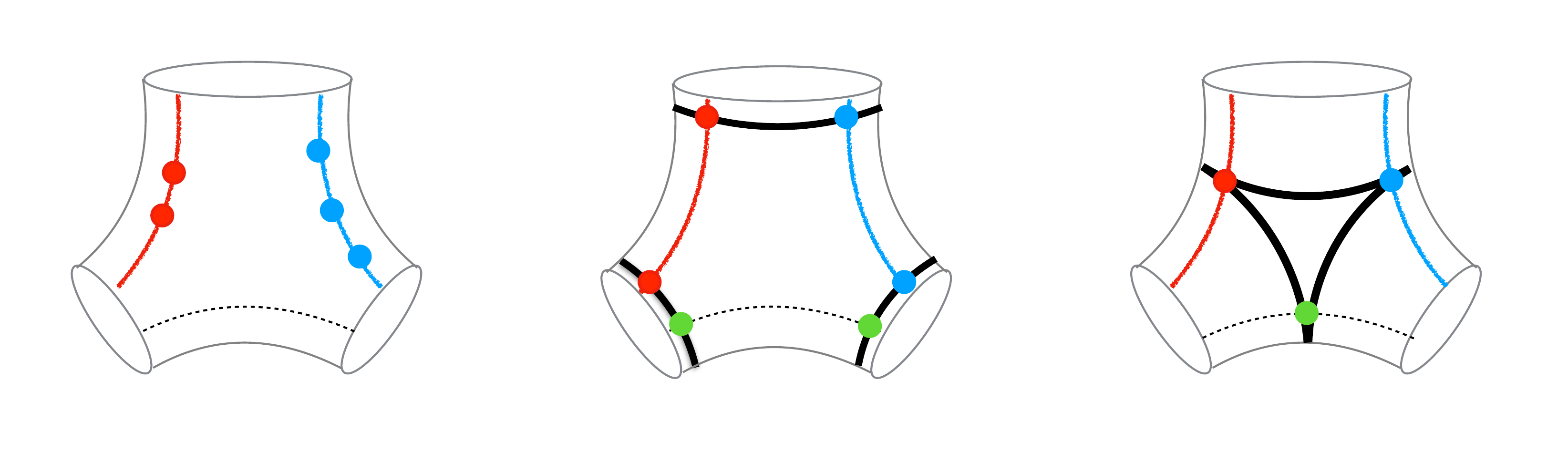}
    \caption{Magnon configurations responsible for different contributions. \textbf{Left:} \textbf{Bridge contribution}. Contributions from magnons on different bridges factorize unless the rapidities of magnons on different bridges coincide, see below. The contribution from the bottom bridge is trivial because there is no twist inserted on that bridge. \textbf{Middle:} \textbf{Wrapping contribution}. When rapidities of magnons on pairs of adjacent bridges coincide, they give rise to contact terms leading to the wrapping contributions, represented by thick black lines. \textbf{Right:} \textbf{Bridge-like contribution}. When magnons living on three different bridges have coinciding rapidities, three-magnon contact terms arise that lead to the bridge-like contribution. }
    \label{diff contributions}
\end{figure*}

To understand the structure of the result, it is convenient to decompose \eqref{Cfact} as $C_{k,\ell,p}^{(\alpha_1,\alpha_2,\alpha_3)}=(\textbf{bridge})\times (\textbf{bridge-like})\times (\textbf{wrapping})$ where
\begin{align}
\label{contributions}
\begin{aligned}
    (\textbf{bridge})&=\det \left(1-s_{\alpha_1} K_{k}\right)\det \left(1-s_{\alpha_2} K_{\ell}\right)\,,\\
    (\textbf{bridge-like})&=\det \left(1-s_{\alpha_3} K_{p}\right)\,,
\end{aligned}
\end{align}
while $(\textbf{wrapping})$ denotes the remaining factors in \eqref{Cfact}, cf. \eqref{bbSn2}. As we will see below, each of them comes from different configuratons of magnons, see Figure \ref{diff contributions}:
\begin{itemize}
    \item Bridge contributions come from magnons on a single bridge.
    \item Wrapping contributions come from contact terms among magnons on two adjacent bridges.
    \item Bridge-like contributions come from contact terms involving magnons on all the three bridges. They have the functional form a bridge contribution \eqref{contributions}, although there is no bridge of the corresponding length.
\end{itemize}
    
Below we sketch the computation of each contribution.
    
\smallskip
    
\noindent {\it Bridge contribution.}  Bridge contributions come from  summing over all possible numbers of magnons on a single bridge and integrating over their rapidities. Doing this for the bridge of length $L$ and twist $\alpha$ gives
\begin{align}
\label{bbBn}
    B^{(\alpha)}_{L} = 1\!+\!\sum_{n=1}^{\infty} \frac{1}{n!} \prod_{k=1}^n\! \left(\sum_{a_k=1}^{\infty}\int\! \frac{\dd u_k}{2\pi}\, e^{-{L} \widetilde{E}_{a_k}(u_k)}\!\right)\! \bbB^{(\alpha)}_n
\end{align}
with 
\begin{align}
\label{BH}
    \bbB^{(\alpha)}_n\equiv\prod_{k=1}^n \Big(\tmu_{a_k}(u_k)\, T^{(\alpha)}_{a_k}\Big) \prod_{i<j}H_{a_i,a_j}(u_i,u_j)\, .
\end{align}
The quantities above depend on the rapidities $u_k$ via the Zhukovsky transform defined by $x+1/x= u/g$. $x_k = x(u_k)$ has a quadratic branch cut from $-2g$ to $2g$, and we denote $x_k^{[\pm a]} = x(u_k\pm \ii{a}/{2})$ for positive integers $a$. Mirror kinematics, usually denoted by $u^\gamma$, corresponds to the analytic continuation $x^{[+a]}(u^\gamma) = 1/x^{[+a]}(u)$ and $x^{[-a]}(u^\gamma) =x^{[-a]}(u)$. The physical momentum $p_a(u)$ and the mirror energy $\tE_a(u)$ are given by
\begin{equation}
    e^{\ii p_a} = {x^{[+a]}}/{x^{[-a]}}\, ,\quad e^{\tE_a} = {x^{[+a]}x^{[-a]}}
\end{equation}
while the measure $ \tmu_a(u)$ and the symmetric hexagon weight $\tH_{ab}(u,v)$ in the mirror kinematics are given by
\begin{align}
\label{measureH}
    \tmu_a=&\frac{1}{\ii g}\prod_{\epsilon =\pm}\frac{1}{x^{[\epsilon a]}- 1/x^{[\epsilon a] }}\,H_a \, ,\\ \no
    \tH_{ab}(u_i,u_j)&=\!\!\prod _{\epsilon, \delta=\pm}\frac{x_i^{[\epsilon a]}-x_j^{[\delta b]}}{x_i^{[\epsilon a]}\, x_j^{[\delta b]}-1}\, ,\ \ H_a\!=\!\frac{x^{[+a]} - x^{[-a]}}{x^{[+a]} x^{[-a]} - 1} .
\end{align}
    
The factor $T^{(\alpha)}_a$ is the character of the twist of the corresponding bridge, $\tau_a^\alpha$, in the $a$-th antisymmetric representation of PSU$(2|2)$ 
\begin{align}
\label{charactertwist}
    T^{(\alpha)}_a=\STr_a \, \tau_a^\alpha=4a s_{\alpha}\, ,
\end{align}
where we define the super-trace with a minus sign for bosonic states.
    
The partition function $B^{(\alpha)}_L$ takes the form of the so-called octagon \cite{Fleury:2016ykk,Coronado:2018ypq}. The sum over the bound states, labeled by $a$, can be explicitly performed, leading to the the weight $\chi_g(t)$ in the Fourier representation  
\begin{equation}
    4 s_\alpha \chi_g(t) = \sum_{a\geqslant 1} T^{(\alpha)}_a \, e^{ta/2g}\, ,
\end{equation}
which allows \eqref{bbBn} to be rewritten \cite{Kostov:2019auq,Kostov:2019stn,Belitsky:2020qir} in the form
\begin{align}
\label{BridgeOctagon}
    B^{(\alpha)}_L = \det (1-s_\alpha K_{L})\, .
\end{align}
Note that $s_0 = 0$, so that this contribution is trivial for the bridge without twist.
\smallskip
    
\noindent {\it Wrapping contribution.}
The hexagon form factors develop poles when the rapidities of magnons in adjacent bridges coincide. This property, known as \textit{decoupling condition}, corresponds physically to a ``magnon-antimagnon'' pair -- or \textit{wrapping magnon} -- decoupling from the hexagon and going to infinity, {\it i.e.} approaching the operator insertion points. The regularization of these singularities using the genus-two surface, discussed later in the text and in more detail in the Supplemental Material, allows to collect contact terms into a partition function of the wrapping magnons. For a given operator that contains the twists $\alpha$ and $\beta$ and has length $L$, we find that the contribution of these contact terms is given by $W^{(\alpha+\beta)}_{L}$, with
\begin{align}
\label{bbSn}
    \left(W^{(\alpha)}_{L}\right)^2\!\! = 1+\sum_{n=1}^{\infty}\frac{1}{n!} \prod_{k=1}^n \! \left(\sum_{a_k=1}^{\infty}\int \frac{\dd u_k}{2\pi} e^{-L \widetilde{E}_{a_k}\!(u)}\right) \mathbb{W}^{(\alpha)}_n  \,,
\end{align}
where
\begin{align}
\label{supetr}
    \mathbb{W}^{(\alpha+\beta)}_n\! &\equiv (-\ii)^n\partial_{\mathbf{v}}\operatorname{STr}\! \left[\tau^{\alpha}_{\mathbf{a}} \,\mathcal{S}_{\mathbf{a}\mathbf{b}}(\mathbf{u},\mathbf{v}) \,\tau^{\beta}_{\mathbf{b}} \,\mathcal{S}_{\mathbf{a}\mathbf{b}}(\mathbf{u},\mathbf{v})\right]\!\Big|_{\,{\bf v}\to {\bf u}}
\end{align}
and
\begin{align}
    \tau_{\mathbf{a}}^{\alpha} = \prod_{k=1}^n \! \tau^{\alpha}_{a_k} \, , \,\,\,\,\mathcal{S}_{\mathbf{a}\mathbf{b}}(\mathbf{u},\mathbf{v}) = \prod_{i=1}^n\! \prod_{j=1}^n \mathcal{S}_{a_i,b_j}(u_i,v_j)\, .
\end{align}
    
Here $\mathcal{S}_{ab}(u,v)$ denotes Beisert's scattering matrix \cite{Beisert:2005tm,Beisert:2006qh,Arutyunov:2009mi} for mirror bound states. After taking the derivatives with respect to $ {\bf v}$, the two groups of $n$ rapidities ${\bf u}$ and $ {\bf v}$ are identified, and then the integrals over ${\bf u}$ are performed. The expression \eqref{supetr} for the all-loop, all-magnon wrapping contributions is one of our main new results. A similar contribution arises for hexagons in the fishnet theory, but in that situation there are many more contributions that do not vanish \footnote{G. Ferrando and E. Olivucci, unpublished.}. Importantly, the result only depends on the product of the twists and sum of the lengths of each bridge. Moreover, for untwisted operators, this contribution is absent. 
    
Although we did not manage to evaluate \eqref{supetr} explicitly for arbitrary $n$, we conjecture that the result matches
\begin{align}\label{bbSn2}
    \left(W^{(\alpha)}_L\right)^2 = \frac{1}{\det (1-s_{\alpha}K_{L-1})\,\det (1-s_{\alpha}K_{L+1})}\, .
\end{align}
We verified this up to three virtual magnons, see the discussions around \eqref{BtoW} for more details. 

\smallskip
    
\noindent {\it Bridge-like contribution.} The most nontrivial contribution is the bridge-like contribution, coming from contact terms in which the rapidities of magnons on three different bridges coincide. They can be also computed using the genus-two regularization. Our analysis, detailed in the Supplemental Material, shows that the three magnons effectively merge, creating an excitation resembling the usual bridge magnons, with the twist given by $\tau^{\alpha_1} \tau^{\alpha_2} = \tau^{\alpha_3}$ and an effective bridge length $\ell+k = p$. The computation of the first few terms in their expansion suggests that 
\begin{align}
    B_{p}^{(\alpha_3)} = \det (1-s_{\alpha_3}K_{p})\,.
\end{align}

\noindent\textbf{Final result.}
As we explain in the Supplemental Material, these contributions factorize, leading to an expression that is simply a product of all the terms. The final answer matches the localization result \eqref{Cfact}. In the rest of this paper, we give more technical details. 
    
\medskip
    
\noindent{\bf  Regularization on the genus-two surface.} As mentioned earlier, we need to address the singularities of the integrand that arise when the rapidities of magnons on different bridges coincide. The importance of such singularities was highlighted in \cite{Basso:2017muf}, where the one-magnon contribution was explicitly evaluated. As shown there, the divergences stem from the infinite size of the mirror channels, where the virtual magnons live. 

\begin{figure*}[t]
    \centering
    \includegraphics[scale=0.5]{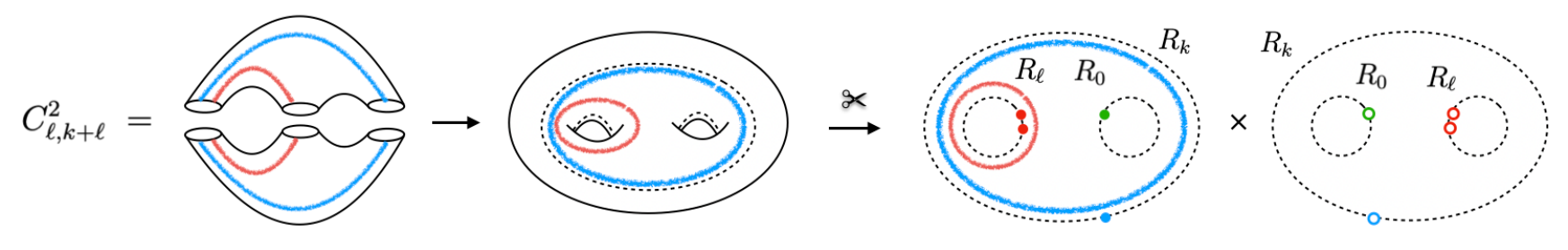}
    \caption{Gluing two three-point functions into a genus-two closed surface, then cutting the result on three mirror (dotted) lines to get two different pairs of pants. Each mirror cut reveals an arbitrary number of magnons  (red, green and blue dots). The lengths $R_\ell,\,R_0,\,R_k$ are supposed to be large so that the physical excitations are suppressed. The twist line is inserted along the mirror bridge of length $R_\ell$. }\label{fig:SquareC} 
\end{figure*}

A natural way to regularize these divergences is to introduce a cut-off on the volume of these spaces. In general, the regularized singularities yield both a divergent, or volume-dependent, term proportional to the anomalous dimensions of the operators, and a finite term contributing to the structure constant. In our case, the divergent term is absent, as the dimension of our operators is protected but the finite term remains nontrivial.
    
A systematic evaluation of such finite  contributions for arbitrary excited operators is still an open problem in the hexagon program for correlation functions in ${\cal N}=4$ SYM theory. In \cite{Basso_IGST22,Basso:2022nny}, it was suggested that both the volume-dependent and finite terms for the three-point function can be controlled by considering the OPE limit of a four-point function. This can be worked out explicitly in the fishnet theory \cite{Gurdogan:2015csr}, where the hexagon approach to correlation functions can be derived from first principles by constructing separated variables for a non-compact, integrable spin chain \cite{Derkachov:2019tzo,Derkachov:2020zvv}. 
    
Here, we take a similar but slightly different approach to \cite{Basso_IGST22,Basso:2022nny} and regulate the infinite volumes by gluing two three-point functions into a genus-two surface. The genus-two surface provides a natural and physical cut-off for the square of the structure constant, whose partition function can be computed by gluing together two pairs of pants with \textit{mirror} edges, see Figure \ref{fig:SquareC}. 
    
We develop this idea in the Supplemental Material and show how the wrapping and bridge-like magnons arise from the contact terms of two and three magnons. We also explain how the factorization happens and give some examples.
\medskip
    
\noindent{\textbf{Structure of the wrapping contribution.}} 
Our conjectured finite-coupling expression for the square of the wrapping contribution \eqref{bbSn2} is remarkably similar to its bridge counterpart \eqref{BridgeOctagon}. This suggests that $\mathbb{W}^{(\alpha)}_n$ can be expressed in a closer way to $\mathbb{B}^{(\alpha)}_n$ in \eqref{bbBn}, which would be ideal for checking the conjecture. 
    
To do so, let us first consider the bridge contribution $B_{L}^{(\alpha)}$, given by the octagon, and take its logarithm. As shown in \cite{basso2019,Kostov:2019stn,Kostov:2019auq}, the octagon can be expressed as a Fredholm Pfaffian and its logarithm admits a simple series expansion:
\begin{align}
    \ln B_{L}^{(\alpha)} = \sum_{n=1}^{\infty} \frac{1}{n} \prod_{k=1}^n \sum_{a_k=1}^{\infty}\int \frac{\dd u_k}{2\pi}\, e^{-L \widetilde{E}_{a_k}} \textrm{C}\bbB_n\,,
\end{align}
where $\textrm{C}\bbB_n$ denotes the ``connected part'' whose details can be found in \cite{basso2019,Kostov:2019stn} and are briefly recalled in the Supplemental Material. Equating this with the expansion of $\ln \det (1-K)$ appearing in \eqref{BridgeOctagon}, we obtain an identity
\begin{align}\label{eq:identityuseful}
    \prod_{k=1}^n \sum_{a_k=1}^{\infty}\int \frac{\dd u_k}{2\pi}\, e^{-L \widetilde{E}_{a_k}} \textrm{C}\bbB_n=-s_{\alpha}^{n}\,\Tr\, K^n_{L}\,.
\end{align}
    
Now, by computing the first few terms of \eqref{bbSn}, we found that the logarithm of the wrapping contribution also admits a simple expansion,
\begin{align}
    2\ln W_{L}^{(\alpha)} = \sum_{n=1}^{\infty} \frac{1}{n} \prod_{k=1}^n \sum_{a_k=1}^{\infty}\int \frac{\dd u_k}{2\pi}\, e^{-L \widetilde{E}_{a_k}}\textrm{C}\mathbb{W}_n \,,
\end{align}
with
\begin{align}
\label{BtoW}
\textrm{C}\mathbb{W}_n&=-\left(e^{\sum_{i=1}^n\tE_{a_i}}+e^{-\sum_{i=1}^n\tE_{a_i}}\right)\,\textrm{C}\bbB_n\,,
\end{align}
Thanks to the identity \eqref{eq:identityuseful}, this immediately implies
\begin{align}
\label{logW2}
    \prod_{k=1}^n \sum_{a_k=1}^{\infty}\int \frac{\dd u_k}{2\pi}\, e^{-L \widetilde{E}_{a_k}}
    \textrm{C}\mathbb{W}_n=s_\alpha^{n}\Tr \left( K^n_{L-1}+K^n_{L+1}\right),
\end{align}
which gives our conjecture \eqref{bbSn2} after the exponentiation.
    
We have verified the relation \eqref{BtoW} up to $n=3$ by taking the derivative and traces in \eqref{supetr} with the help of a computer. For $n\leqslant 2$, the verification can be done by hand using the following partial trace identities:
\begin{align}\label{partial traces}
    \STr_a\,\Sm^{-1}_{ab}\, \tau_a^{\alpha}\,\Sm_{ab} &= T_a^{(\alpha)}\,\tH_{ab}\,\textbf{1}_b\, ,\\ \no
    2\ii \, \STr_a\,\Sm^{-1}_{ab}\, \p_u\Sm_{ab} &= \tK_a(u)\, (1-\tH_{ab}) \, \textbf{1}_b\, ,\\ \no
    2\, \STr_{ab}\, \Sm^{-1}_{ab}\, \p_u\p_v\Sm_{ab} &= p'_a(u)\, p'_b(v)\,(1-\tH_{ab})\, ,
\end{align}
where $\tH_{ab}\equiv\tH_{ab}(u,v)$, we define $\tK_a(u)$ and $p'_a(u)$ via
\begin{equation}
    \tK_a\pm p'_a =-2\,e^{\pm\tE_a} \tmu_a  \, ,
\end{equation}
and we recall that $\Sm_{ab}\equiv\Sm_{ab}(u,v)$ is the mirror S-matrix. The indices on the super-trace indicates which spaces are traced over. For higher $n\geqslant 3$ more complicated multiple traces occur. However, for all $n$, the fundamental building blocks remain the same: $p'_a$, $k_a$ and $H_{ab}$. 
    
\medskip
    
\noindent{\textbf{Conclusion and outlook.}} We generalized the hexagon formalism to  the $\mathbb{Z}_K$ orbifold $\mathcal{N}=2$ SCFT, obtaining closed form expressions of various building blocks and reproducing localization results up to three virtual magnons (but nonperturbative in the 't Hooft coupling). A general proof seems within reach. Our findings suggest that recent progress in integrability for $\mathcal{N}=4$ SYM can be extended to a broader class of integrable $\mathcal{N}=2$ SCFTs, opening numerous future directions: 
\begin{itemize}
    \item Studying the long-quiver limit of the $\mathbb{Z}_K$ orbifold quiver $\mathcal{N}=2$ SCFTs,  discussed in e.g.~\cite{Mukhi:2002ck,Korchemsky:2025eyc}. 
    \item Applying this method to other correlation functions in $\mathbb{Z}_K$-orbifold theories, e.g.~three-point functions of non-BPS operators and higher-point functions.
    \item Investigating whether planar three-point functions away from the $\mathbb{Z}_K$-orbifold points can also be expressed as Fredholm determinants, which may hint at integrability beyond the orbifold point \cite{Gadde:2010zi,Pomoni:2021pbj, Bertle:2024djm, Bozkurt:2024tpz,Bozkurt:2025exl}.

    \item Applying this method to regularize the three-point functions to other theories obtained by twisting $\mathcal{N}=4$ SYM, see related works \cite{Basso:2018cvy,Eden:2022ipm}, and in particular to theories in which the vacuum states acquire anomalous dimensions. Such a framework would help further the links between three-point functions in the hexagon formalism and the Quantum Spectral Curve approach \cite{Gromov:2013pga}, along the lines provided in \cite{Basso:2022nny}.
    
    \item Given the simplicity of the localization result, it is worth exploring a more efficient integrability framework beyond the hexagon formalism. One possibility is to directly bootstrap the decompactified string vertex \cite{Bajnok:2015hla,Bajnok:2015ftj,Bajnok:2017mdf}.

    \item There is an alternative localizaton expression for the structure constants, given by a derivative of the two-point function with respect to the 't Hooft coupling \cite{Beccaria:2021hvt, Beccaria:2022ypy,  Billo:2022xas, Billo:2022lrv, Billo:2022fnb, Billo:2022gmq}. Its structure is reminiscent of the stratification for non-planar correlation function \cite{Bargheer:2017nne,Bargheer:2018jvq}, which is relevant for graphs with zero-length bridges. Also in this case, the extremality of three-point functions leads to a zero-length bridge and it would be interesting to understand a potential connection.
\end{itemize}
    
\medskip

\noindent{\textbf{Acknowledgements.}} We are grateful to B.~Basso, M.~Billò, L.~Dixon, S.~Ekhammar, A.~Georgoudis, N.~Gromov,  G.~Korchemsky, E.~Olivucci, E.~Pomoni and A.~Testa for insightful discussions. GL and DS thank the CNRS and ANR for support via the IRP project NP-Strong and ANR project Observables (ANR-24-CE31-7996), respectively. The work of GF is funded by the Deutsche Forschungsgemeinschaft (DFG, German Research Foundation) – Projektnummer 508889767. GL acknowledges support from the European Cooperation in Science and Technology (COST) Action CA22113 “Fundamental challenges in theoretical physics” (THEORY-CHALLENGES) through a Short-Term Scientific Mission grant. We gratefully acknowledge support from CERN, ICTP-SAIFR Sao Paulo, PI Waterloo and Simons Center for Geometry and Physics, Stony Brook University, where various stages of this project were completed.

\bibliography{MyBib}

\onecolumngrid
\clearpage

\section*{Supplemental Material}
    
\setcounter{subsection}{0}
    
\subsection*{Useful properties} 
    
We collect here some properties of the dynamical factors and S-matrix for arbitrary bound states that will be used below, following \cite{Basso:2015zoa,Basso:2018cvy},
\begin{equation}\label{huvproperties}
    h_{ab}(u^{2\gamma},v^{2\gamma}) = h_{ab}(u,v)\, ,\quad h_{ab}(u^{4\gamma},v) = \frac{1}{h_{ba}(v,u)}\, ,
\end{equation}
\begin{equation}
    h_{ab}(u,v) \,h_{ba}(v,u) = H_{ab}(u,v)\, ,\quad H_{ab}(u^{2\gamma},v) = \frac{1}{H_{ab}(u,v)}\, ,
\end{equation}
\begin{equation}\label{hresidue}
    h_{aa}(u^{2\gamma},u) = 1\, ,\quad \lim_{v\to u} \,\frac{(v-u)}{h_{a b}(u,v)} = \frac{\ii\,\delta_{a b}}{\mu_a(u)}\, ,
\end{equation}
\begin{equation}\label{gammasS}
    \mathcal{S}_{ab}(u^{2\gamma},v^{2\gamma}) = \mathcal{S}_{ab}(u,v)\, ,\quad \mathcal{S}_{ab}(u^{4\gamma},v) = \mathcal{S}_{ab}(u,v^{4\gamma}) = \kappa_a\, \mathcal{S}_{ab}(u,v)\, \kappa_a = \kappa_b\, \mathcal{S}_{ab}(u,v)\, \kappa_b\, ,
\end{equation}
with the diagonal matrix $\kappa_a$ being defined below. We also have 
\begin{equation}
    \mathcal{S}_{aa}(u,u) = \mathcal{P}^{g}\, ,
\end{equation}
where $\mathcal{P}^g$ is the graded permutation, which acts naturally on the S-matrices according to $\mathcal{P}_{12}^{g} \mathcal{S}_{23}= \mathcal{S}_{13} \mathcal{P}_{12}^{g}$. The unitarity of the S-matrix then reads
\begin{equation}\label{inverse}
    \mathcal{S}^{-1}_{12}(u,v) =  \mathcal{S}_{21}(v,u) = \mathcal{P}_{12}^{g} \mathcal{S}_{12}(u,v) \mathcal{P}_{12}^{g}\, .
\end{equation}
The crossing relation is given by
\begin{equation}
\label{CrossingS}
    \mathcal{S}_{ab}(u^{2\gamma},v) = \frac{1}{h_{ab}(u,v)\, h_{ab}(u^{2\gamma},v)} \,\mathcal{C}_a \, {}^{t_a}(\mathcal{S}_{ab}^{-1})(u,v) \,\mathcal{C}^{-1}_a\, ,
\end{equation}
where the crossing matrix satisfies ${}^{t_a}\mathcal{C}_a = (-1)^a\mathcal{C}_a$, $\mathcal{C}_a^2 = \kappa_a = \text{diag}(-1_B,1_F)$, $\mathcal{C}_a {}^{t_a}\mathcal{S}_{ab} \mathcal{C}^{-1}_a = \mathcal{C}_b {}^{t_b}\mathcal{S}_{ab} \mathcal{C}^{-1}_b$. The superscript ${}^{t_a}$ indicates the partial super-transposition in the space $a$. The super-transpose of an arbitrary matrix $M$ is given in components by
\begin{equation}
    ({}^{t}\! M)_{ij} = (-1)^{\mf_i \mf_j + \mf_i} M_{ji}\, ,
\end{equation}
where $\mf_i$ is the fermion number in state $i$. Notice that this definition implies that 
\begin{equation}\label{trace graded transpose}
    {}^{t_a}({}^{t_a}M) = \kappa_a M \kappa_a\, , \qquad\Tr_a[{}^{t_a}\!M\, {}^{t_a}\!N] = \STr_a[M \kappa_a N]\, ,
\end{equation}
for arbitrary matrices $M$ and $N$. From the crossing relation and equation \eqref{gammasS}, one deduces the ``crossed unitarity'' property
\begin{equation}\label{inverse partial transpose}
    {}^{t_a} \Sm_{ab}\,{}^{t_a}\! \left(\Sm_{ab}^{-1}\right) = \tH_{ab}\, \textbf{1}_{ab}\, .
\end{equation}
Using this relation and the observation \eqref{trace graded transpose}, we get that
\begin{equation}\label{partial trace M}
    \STr_a\,\Sm_{ab}\,M_a\,\Sm^{-1}_{ab} = \STr_a\,\Sm^{-1}_{ab}\, M_a\,\Sm_{ab} =\STr_a(M_a)\,\tH_{ab}\,\textbf{1}_b\, ,
\end{equation}
where the arbitrary matrix $M_a$ does not act on the space $b$. The following two particular cases are especially useful:
\begin{equation}\label{partial trace tau}
    \STr_a\,\Sm_{ab}\,\tau_a\,\Sm^{-1}_{ab} = \STr_a\,\Sm^{-1}_{ab}\, \tau_a\,\Sm_{ab} =\STr_a(\tau_a)\,\tH_{ab}\,\textbf{1}_b\, .
\end{equation}
where $\tau$ can be any of our twist matrices, and
\begin{equation}\label{partial trace perm}
    \STr_a\,\Sm_{ab}\,\mathcal{P}^g_{ac}\, \Sm^{-1}_{ab} = \STr_a\,\Sm^{-1}_{ab}\, \mathcal{P}^g_{ac}\, \Sm_{ab} = \tH_{ab}\,\textbf{1}_{bc}\, .
\end{equation}
Mathematica experiments with the code provided in \cite{DeLeeuw:2020ifb} suggest that the following formulas also hold,
\begin{align}\label{partial traces derivatives}
    2\ii \, \STr_a\,\Sm^{-1}_{ab}\, \p_u\Sm_{ab} &= \tK_a(u)\, (1-\tH_{ab}) \, \textbf{1}_b\, ,\\
\label{traces derivatives}   
    2\, \STr_{ab}\, \Sm^{-1}_{ab}\, \p_u\p_v\Sm_{ab} &= p'_a(u)\, p'_b(v)\,(1-\tH_{ab})\, ,
\end{align}
with $p'_a$ and $k_a$ defined through
\begin{equation}\label{pkdef}
    \tK_a\pm p'_a =-2\,e^{\pm\tE_a} \tmu_a  \, , \qquad p_a=\ii\ln( x^{[-a]}/x^{[+a]})\,.
\end{equation}
We leave the analytical proof of these formulas for a subsequent work.

\subsection*{Twists and $\mathbb{Z}_K$ orbifolds}
    
As discussed in the main text, the theory obtained by the $\mathbb{Z}_K$ orbifolding of $\mathcal{N}=4$ SYM with gauge group $SU(N K)$  has $K$ distinct operators of length $L$ of the type
\begin{align} \label{eq:twistedSector}
    \mathcal{O}^{(\alpha)}_{L} = \Tr\left(\gamma^\alpha Z^L\right) \text{,} \quad 0\leqslant \alpha\leqslant K-1\, ,
\end{align}
where $\gamma\, Z \,\gamma^{-1}=Z$ with $\gamma = \left(\mathbf{1}_N, \rho \  \mathbf{1}_N ,\dots , \rho^{K-1} \ \mathbf{1}_N\right)$ and  $\rho = e^{2 \pi \ii/K}$. These operators are associated with vacua in the spin-chain representation. The operator $\mathcal{O}_L^{(0)}$ is called ``untwisted'' and $\mathcal{O}_L^{(\alpha)}$ with $\alpha=1,\ldots, K-1$ are called ``twisted''. The action of $\gamma$ on the other scalar fields of the original theory is:
\begin{align}
    \gamma\, (X,Y)\,\gamma^{-1} &= \rho \,(X,Y)\,, \\  \gamma \,(\bar{X},\bar{Y})\,\gamma^{-1} &= \rho^{-1} (\bar{X},\bar{Y})\,,
\end{align}
while the covariant derivatives are left untouched and some fermions are twisted. These fields represent magnons on top of the vacua and can be written in the bi-fundamental representation of PSU(2|2)$^2$ with the magnons $(\dot{\varphi}_1,\dot{\varphi}_2,\dot{\psi}_1,\dot{\psi}_2)\times(\varphi_1,\varphi_2,\psi_1,\psi_2)$,
\begin{align}
    X \sim \dot{\varphi}_1\varphi_1 , \ \ \ \ \bar{X} \sim - \dot{\varphi}_2 \varphi_2, \ \ \ \ Y \sim \dot{\varphi}_2 \varphi_1 , \ \ \ \ \bar{Y} \sim  \dot{\varphi}_1 \varphi_2, \ \ \ \ \mathcal{D}^{\beta \dot{\beta}} \sim  \dot{\psi}^{\dot{\beta}} \psi^\beta\,.
\end{align}
    
One can then represent the effect of conjugation by $\gamma$ by the multiplication $\varphi_1 \rightarrow \rho \  \varphi_1$ and $\varphi_2 \rightarrow \rho^{-1} \  \varphi_2$  while keeping $\dot{\varphi}$, $\psi$ and $\dot{\psi}$ unchanged, that is, acting with a twist $(1,1,1,1)_L \times\tau_R=(1,1,1,1)_L \times (\rho,\rho^{-1},1,1)_R$ on the magnons.  This motivates us to implement the matrices $\gamma^\alpha$ in the hexagon computation as insertions of powers $\tau^\alpha$ of the twist $\tau$. We insert the twists in the legs between chiral $ \mathcal{O}^{(\alpha)}_{L}$ and anti-chiral $ \bar{\mathcal{O}}^{(\alpha)}_{L}$ operators as in Figure \ref{fig:WrappedOctagon}. The twist is breaking the $\textrm{PSU(2|2)}_R$ symmetry down to
$\textrm{SU(2)} \times \textrm{SU(2)} $ for $K=2$ and 
to $\textrm{SU(2)} \times \textrm{U(1)} $ for $K>2$.

\subsection*{Regularization on the genus-two surface}
    
In this section we outline the regularization procedure for the three-point function. As described in the main text, we begin by gluing two pair of pants along three long legs of lengths $r_{k}$, $r_\ell$ and $r_p$ associated to the operators $\mathcal{O}_k^{(\alpha_1)}$, $\mathcal{O}^{(\alpha_2)}_\ell$ and ${\bar{\mathcal{O}}}^{(\alpha_3)}_{p}$, respectively. Next, we compute the resulting genus-two surface by cutting it open into two pairs of pants with mirror magnons, like in Figure \ref{fig:SquareC}. Now, the lengths $R_\ell=r_\ell+r_{p}$, $R_0=r_\ell+r_{k}$, $R_k=r_k+r_{p}$ serve as regulators for the volume of the mirror channels. We have to sum over an arbitrary number of these magnons on each cut. We recover the square of the three-point function in the limit $r_j\to \infty$.

\begin{figure}[t]
      \includegraphics[scale=0.35]{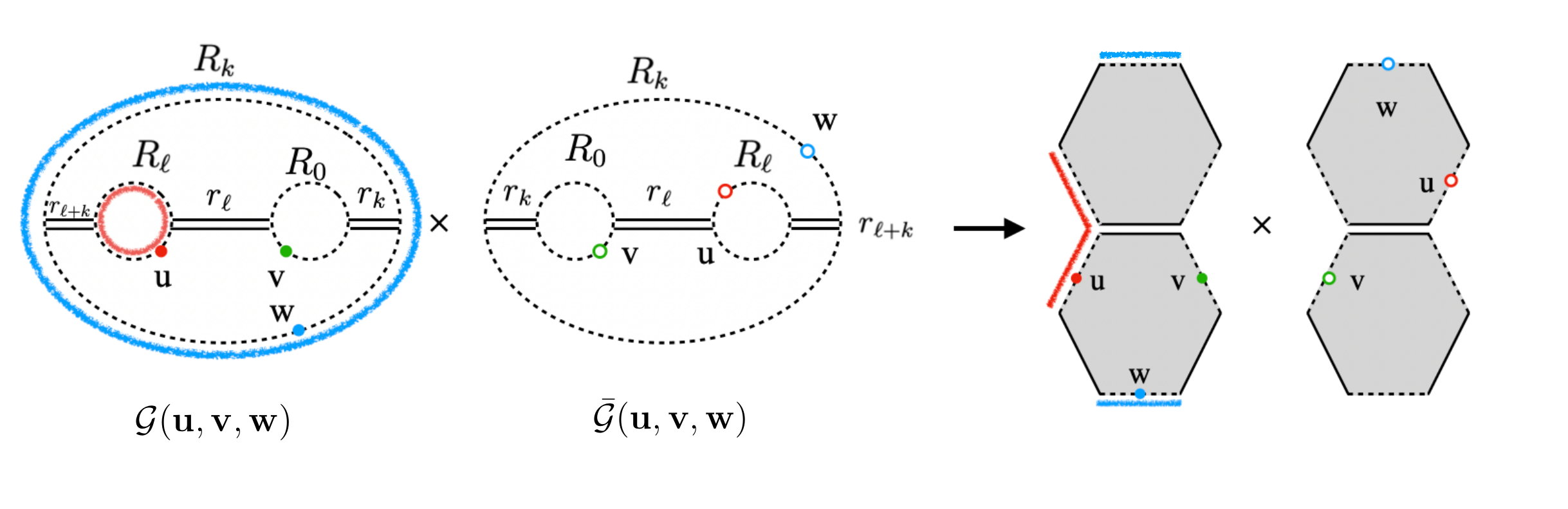}
      \caption{Cutting open the two pairs of pants ${\mathcal{G}}(\u,\v,\w)$ and $\bar {\mathcal{G}}(\u,\v,\w)$ into four hexagons on physical bridges of lengths $r_{\ell+k}$, $r_\ell$ and $r_k$ respectively. Here, the sets of magnons $\u$, $\v$ and $\w$ are represented by a single magnon, for simplicity.}\label{fig:TwoOctagons} 
    \end{figure}

The object we compute can be written schematically as
\begin{align}
\label{Cpants}
    \left(C^{(\alpha_1,\alpha_2,\alpha_3)}_{k,\ell,p}\right)^2 &= \sum_{n_u,n_v,n_w\geqslant 0}{\mathcal{C}_{(n_u,n_v,n_w)}} \\ \no
    \mathcal{C}_{(n_u,n_v,n_w)} &= \frac{1}{n_u!n_v!n_w!}\lim_{r_j\to \infty} \int \dd\u\, \dd\v\, \dd\w\, \mu(\u)\,\mu(\v)\,\mu(\w)\, e^{-\ell \tE(\u)- k\tE(\w)}\,\STr\left[\bf{\tau}^{\alpha_2}_\u  \, {\mathcal{G}}(\u,\v,\w)\,\bf{\tau}^{\alpha_1}_\w\,\overline{\mathcal{G}}(\u,\v,\w)\right]\, ,
\end{align}
where $n_u,n_v,n_w$ denote the numbers of magnons in the sets $\u,\v,\w$. The twists $\tau^{\alpha_2}_\u$ and $\tau^{\alpha_1}_\w$ are inserted for the magnons on the bridge of length $\ell$ and $k$, respectively, while the magnons $\v$ on the bridge of zero length see a trivial twist, $\tau_\v^0$. We can now think of the mirror pair of pants as asymptotic three-point functions, each of which can be cut into two hexagons by splitting the rapidities in all the possible ways between the two hexagons, as shown in Figure \ref{fig:TwoOctagons}.This can be written schematically as
\begin{equation}
\label{OtoHexagons}
    \mathcal{G}(\mathbf{u},\mathbf{v},\mathbf{w}) = \sum_{\substack{\beta_u \cup \bar{\beta}_u = \mathbf{u} \\ \beta_v \cup \bar{\beta}_v = \mathbf{v} \\ \beta_w \cup \bar{\beta}_w = \mathbf{w}}} w_{R_\ell}(\beta_u,\bar{\beta}_u) \, w_{R_0}(\beta_v,\bar{\beta}_v) \, w_{R_k}(\beta_w,\bar{\beta}_w) \, \mathcal{H}(\beta^{4\gamma}_v,\beta^{2\gamma}_w,\beta_u) \,\mathcal{H}(\bar{\beta}^{4\gamma}_u,\bar{\beta}^{2\gamma}_w,\bar{\beta}_v)\, ,
\end{equation}
where $\beta_u = (u_{i_1},\dots,u_{i_{|\beta_u|}})$ and the indices are ordered: $i_1<\dots <i_{|\beta_u|}$. When all the magnons are on the same mirror edge, the hexagon weights $\mathcal{H}(\mathbf{u})$ are given by \cite{Basso:2015zoa,Basso:2018cvy}
\begin{equation}
\label{hexadef}
    \mathcal{H}(\mathbf{u}) = (-1)^{\mathfrak{f}}\left(\prod_{i<j} h(u_i,u_j)\right)\left(\cdots \mathcal{S}_{23} \mathcal{S}_{13} \mathcal{S}_{12}\right)\, ,
\end{equation}
where $\mathfrak{f}$ is a grading factor that we do not need to specify. The crossings $2\gamma$ or $4\gamma$ of the rapidities that appear in \eqref{OtoHexagons} account for moving some of the magnons to other mirror edges. The transition factors $w_{R}$ account for the moves of magnons from one hexagon to the other. They include a phase factor and a product of scattering matrices that represent the magnon reordering necessary for the partitioning. For instance, $w_{R_\ell}(\beta,\bar{\beta})$ is given by
\begin{align}
\label{transitionf}
    w_{R_\ell}(\beta,\bar{\beta}) = (-1)^{|\bar \beta|}\prod_{u_j \in \bar{\beta}}  e^{\ii\, \tilde p(u_j)\, r_\ell}  \prod_{\substack{u_j \in \bar{\beta}, u_k \in \beta \\ j<k }} \mathbb{S}(u_j,u_k) =
    (-1)^{|\bar \beta|}\prod_{u_j \in \bar{\beta}}  e^{-\ii\, \tilde p(u_j)\, r_{\ell+k}}  \prod_{\substack{u_j \in \bar{\beta}, u_k \in \beta \\ j>k }} \mathbb{S}(u_k,u_j)\,, 
\end{align}
with $\mathbb{S}(u,v)={S}_0 (u,v)\,\mathcal{S}(u,v)\otimes \mathcal{S}(u,v)$, with ${S}_0 (u,v)=h(u,v)/h(v,u)$ and $\mathcal{S}(u,v)$ the scalar factor and  Beisert's $\textrm{PSU}(2|2)$ scattering matrix respectively, all in mirror kinematics. Furthermore, $\tilde p_a(u) = g\left(x^{[a]} - \frac{1}{x^{[a]}} + x^{[-a]} - \frac{1}{x^{[-a]}}\right)$ denotes the mirror momentum. Transitioning from one hexagon to the other can be done in two different ways, for example the magnons $\u$ in the left pants in Figure \ref{fig:TwoOctagons} can transition from the upper left hexagon to the lower left hexagon either through the bridge of length $r_\ell$ or through the one of length $r_{\ell+k}$. When $R_\ell=r_\ell+r_{\ell+k}$ is finite, the equivalence of the two type of transitions in \eqref{transitionf} is insured by the Bethe equations. 
    
Below, we will choose either of the two forms of the transition factor depending on what we find more convenient for the large-volume limit, keeping in mind that they are equivalent.

\subsection*{Details of the computation}

In this section we will compute the different contributions of the virtual magnons to the square of the structure constant. Let us first briefly explain the origin of the various factors.
    
\begin{itemize}
    \item {\it Bridge contributions.} Summing over configurations with all magnons on the same bridge yields the squares of the bridge contributions,
    \begin{equation}
    \label{CCBl}
    \sum_{n_u=0}^\infty\mathcal{C}_{(n_u,0,0)} = \left(\!B^{(\alpha_2)}_\ell\right)^2\, ,\qquad  \sum_{n_v=0}^\infty\mathcal{C}_{(0,n_v,0)} = 1\, ,\qquad  \sum_{n_w=0}^\infty\mathcal{C}_{(0,0,n_w)} = \left(\!B^{(\alpha_1)}_k\right)^2\, .
    \end{equation}
    \item {\it Wrapping contributions.} Hexagons have singularities when magnons sitting on two different mirror edges have coinciding rapidities and bound-state indices. Some of these singularities give rise to contact terms which we call wrapping contributions. The easiest way to single them out is to study configurations with the same number of magnons on two different mirror bridges (see Figure \ref{fig:WrappingTerms}); and we claim that
    \begin{equation}
    \sum_{n=0}^\infty\mathcal{C}_{(n,n,0)} = \left(\! W^{(\alpha_2)}_\ell\right)^2 \,,\qquad \sum_{n=0}^\infty \mathcal{C}_{(0,n,n)} = \left(\! W^{(\alpha_1)}_k\right)^2 \, .
    \end{equation}
    More generally, we expect the following partial sums to factorize according to
    \begin{equation}
    \label{CCW}
    \sum_{n_u,n_v=0}^\infty \!\!\!\! \mathcal{C}_{(n_u,n_v,0)} = \! \left( \! W^{(\alpha_2)}_\ell B^{(\alpha_2)}_{\ell} \right)^2\!,\ \  \sum_{n_v,n_w=0}^\infty\!\!\!\! \mathcal{C}_{(0,n_v,n_w)} = \! \left(\! W^{(\alpha_1)}_{k} B^{(\alpha_1)}_{k}\! \right)^2\!,\ \  \sum_{n_w,n_u=0}^\infty\!\!\!\! \mathcal{C}_{(n_u,0,n_w)} = \! \left( \! B^{(\alpha_1)}_{k} B^{(\alpha_2)}_{\ell} W^{(\alpha_3)}_p \right)^2\! .
    \end{equation}
    \item 
    {\it Bridge-like contributions.} The decoupling singularities also give rise to a contribution that looks like a regular bridge contribution with effective bridge length $p$. They come from contact terms where magnons excitations in three different channels coincide. The simplest way to study them is thus to consider configurations with the same number of magnons in all three mirror edges, as in Figure \ref{fig:BridgeLike}.
\end{itemize}

\subsubsection*{Bridge}

\begin{figure}[t]
    \includegraphics[scale=0.5]{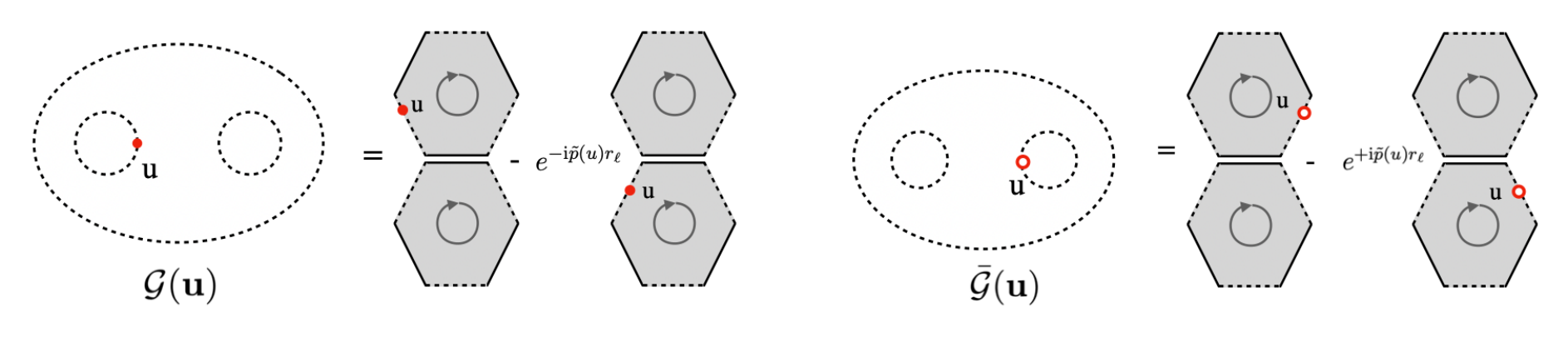}
    \caption{Cutting mirror pants into hexagons along a physical bridge of length $r_\ell$ and transporting excitations from one hexagon to another. An orientation is needed to define the transport factors for the magnons moving from a hexagon to the other. A magnon with momentum $\tilde p(u)$ gets a factor $e^{\ii\tilde p(u)r_\ell}$ when it crosses the bridge of length $r_\ell$ in the direction of the arrow, as in right pair of pants, and $e^{-\ii\tilde p(u)r_\ell}$ when it moves against them, as for the left one. }
    \label{fig:OctagonsToHexagons}
\end{figure}

{\it One-magnon bridge contribution.} Let us start by considering the configuration with one single magnon $u$ in the mirror bridge with length $R_\ell$ gluing the two mirror pair of pants in Figure \ref{fig:SquareC}. Using \eqref{Cpants}, this is equal to
\begin{align}
\label{one bridge magnon}
    \mathcal{C}_{(1,0,0)} = \lim_{r_\ell\rightarrow \infty} \sum_{a=1}^\infty \int \frac{\dd u}{2\pi} \mu_a(u) e^{-\ell \tE_a(u)} \STr_a \left[\tau^{\alpha_2}_a \,\mathcal{G}(u,\emptyset,\emptyset)\, \overline{\mathcal{G}}(u,\emptyset,\emptyset)\right]
\end{align}
with $\mathcal{G}(u,\emptyset,\emptyset)$ and $\overline{\mathcal{G}}(u,\emptyset,\emptyset)$ represented in Figure \ref{fig:OctagonsToHexagons}. The transition factor $e^{-\ii\tilde p_a(u)r_\ell}$ appears when a magnon crosses from the upper to the lower hexagon on the left pair of pants $\mathcal{G}(u,\emptyset,\emptyset)$, while $e^{\ii\tilde p_a(u)r_\ell}$ appears when a magnon crosses from the upper to the lower hexagon on the right pair of pants, $\overline{\mathcal{G}}(u,\emptyset,\emptyset)$. 

Plugging these building blocks in \eqref{one bridge magnon}, we get
\begin{equation}
\label{nl0k}
    \mathcal{C}_{(1,0,0)} = \lim_{r_\ell\to \infty} \sum_{a=1}^\infty\int \frac{\dd u}{2\pi}\,  \mu_a(u)\, e^{-\ell \tE_a(u)} \,T_a^{(\alpha_2)} \left(1-e^{-\ii\tilde p_a(u)r_\ell}-e^{\ii\tilde p_a(u)r_\ell}+1\right) = 2\sum_{a=1}^\infty\int \frac{\dd u}{2\pi}\,e^{-\ell \tE_a(u)}\, \mathbb{B}^{(\alpha_2)}_1
\end{equation}
with $T_a^{(\alpha)}=\STr\, \tau^{\alpha}_a=4a s_\alpha$. When $r$ is large, the two middle terms in \eqref{nl0k} oscillate rapidly and their contributions are thus suppressed when integrated. Notice that we get \textit{twice} the one-magnon bridge contribution, as expected since we are computing the square of the structure constant.
    
{\it Multi-magnon bridge contributions.}
This argument extends almost straightforwardly to $\mathcal{C}_{(n,0,0)}$, with $n$ excitations on one bridge and none on the other two.
\begin{align}
    \mathcal{C}_{(n,0,0)} = &\lim_{r\rightarrow \infty}\frac{1}{n!} \int \dd\u \,\mu(\u)\,e^{-\ell \tE(\u)} \sum_{\substack{\alpha \cup \bar{\alpha} = \mathbf{u} \\ \beta \cup \bar{\beta} = \mathbf{u}}}(-1)^{|\bar{\alpha}|+|\bar{\beta}|} \, e^{-\ii (\tilde{p}(\bar{\alpha})-\tilde{p}(\bar{\beta})) r_\ell}\, h_<(\alpha) h_<(\bar{\alpha})\, h_>(\beta) h_>(\bar{\beta})  \cdots\, ,
\end{align}
where
\begin{equation}h_<(\beta) = \prod_{\substack{u_j,u_k \in \beta \\ {j<k}}}h_{a_j,a_k}(u_j,u_k)\,,\qquad h_>(\beta) = \prod_{\substack{u_j,u_k \in \beta \\ {j>k}}}h_{a_j,a_k}(u_j,u_k)\,,\qquad H(\beta)=h_<(\beta)\, h_>(\beta)\,,
\end{equation}
and the dots stand for a super-trace of a product of  S-matrices that  contains no decoupling poles. In the limit $r_\ell\to\infty$, the only terms that survive are those where there are no exponential factors left. This corresponds to situations where the left and right octagons are mirror images of one another, {\it i.e.} $\alpha = \beta$. In those cases, the S-matrices simply cancel and the super-trace trivially reduces to $\STr\,\prod_{k=1}^n \tau^{\alpha_2}_{a_k} = \prod_{k=1}^n T^{(\alpha_2)}_{a_k}$. The sum then becomes 
\begin{align}
    \mathcal{C}_{(n,0,0)} &= \frac{1}{n!} \prod_{k=1}^n \left(\sum_{a_k=1}^{\infty} \int \frac{\dd u_k}{2\pi}\, \mu_{a_k}(u_k)\,e^{-\ell \widetilde{E}_{a_k}(u_k)}\, T^{(\alpha_2)}_{a_k}\right)  \sum_{\beta \cup \bar{\beta} = \mathbf{u}} H(\beta)\, H(\bar{\beta})\\
    &= \frac{1}{n!} \prod_{k=1}^n \left(\sum_{a_k=1}^{\infty} \int \frac{\dd u_k}{2\pi}e^{-\ell \widetilde{E}_{a_k}(u_k)}\right) \sum_{m=0}^n \binom{n}{m}\, \bbB^{(\alpha_2)}_m \,\bbB^{(\alpha_2)}_{n-m}\, ,
\end{align}
which we recognize as the $n$-th order term in the expansion of $\left(\!B^{(\alpha_2)}_\ell\right)^2$. Therefore,
\begin{equation}
    \sum_{n=0}^\infty\mathcal{C}_{(n,0,0)} = \left(\!B^{(\alpha_2)}_\ell\right)^2\, .
\end{equation}
The situation is obviously similar for the other bridges, and we obtain $\eqref{CCBl}$.

\subsubsection*{Wrapping}
    
\begin{figure}[t]
\includegraphics[scale=0.50]{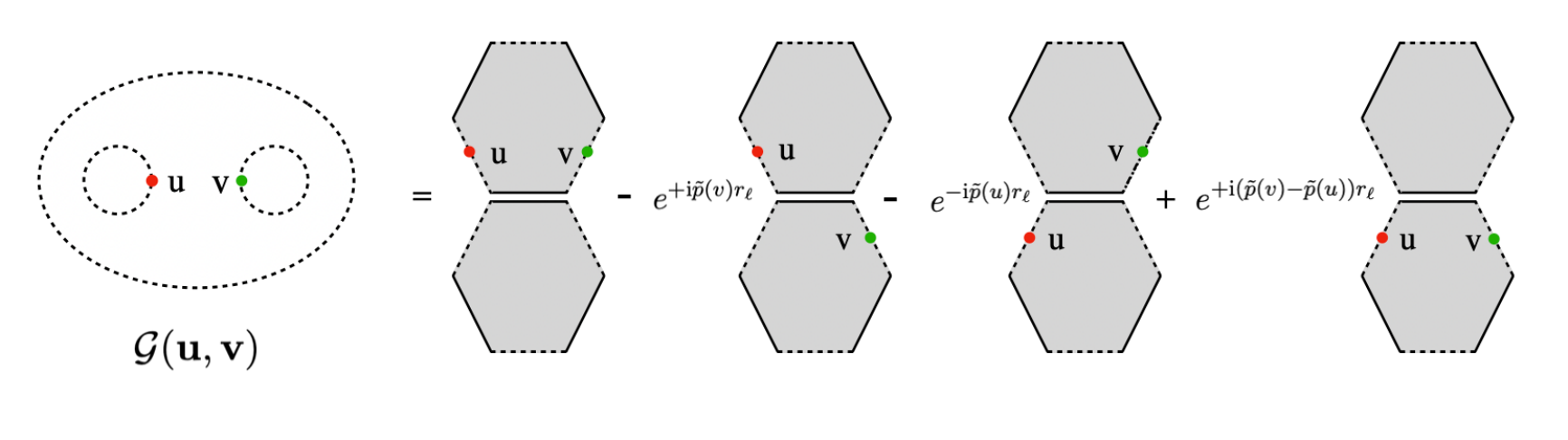}
\caption{Distribution of two magnons on opposite sides of the first pair of pants, according to \eqref{contactO1}.}
\label{fig:TwoMagnonsLH}
\end{figure}

{\it One-wrapping contribution.} As was explained before, the decoupling poles are responsible for the wrapping contributions. The simplest configurations for which they appear involve two magnons in different mirror bridges, such as $\mathcal{C}_{(1,1,0)}$ where one magnon is on the bridge of length $\ell$ and the other that of length $0$. In the following, we describe how the regularization of these poles generates the wrapping contributions. According to our prescription, we begin by cutting the pairs of pants $\mathcal{G}(u,v,\emptyset)$ and  $\bar{\mathcal{G}}(u,v,\emptyset)$ into hexagons, as shown in Figure \ref{fig:TwoMagnonsLH},
\begin{align}
\label{contactO1} 
    \mathcal{G}(u,v,\emptyset)= \frac{\kappa_a\,\mathcal{S}_{ba}(v,u)\,\kappa_a}{h_{ab}(u,v)} - e^{ \ii \tilde{p}_b(v)r_\ell} - e^{-\ii \tilde{p}_a(u)r_\ell} + e^{\ii(\tilde{p}_b(v) - \tilde{p}_a(u))r_\ell} \,\frac{\kappa_a\,\mathcal{S}_{ab}(u,v)\,\kappa_a}{h_{ba}(v,u)}\, , \\
\label{contactO2}
    \bar{\mathcal{G}}(u,v,\emptyset)= \frac{\kappa_a\,\mathcal{S}_{ab}(u,v)\,\kappa_a}{h_{ba}(v,u)} - e^{-\ii \tilde{p}_b(v) r_\ell} - e^{\ii\tilde{p}_a(u) r_\ell} + e^{\ii (\tilde{p}_a(u) - \tilde{p}_b(v))r_\ell} \,\frac{\kappa_a\,\mathcal{S}_{ba}(v,u)\,\kappa_a}{h_{ab}(u,v)}\, ,
\end{align}
where the $\kappa_a$ factors come from using the property \eqref{gammasS} of the S-matrix.
    
The next step is to compute  $\STr_{ab}\left[ \tau^{\alpha_2}_a  \,\CG(u,v,\emptyset)\, \bar\CG(u,v,\emptyset)\right]$. Taking the trace with the help of \eqref{partial trace tau}, we see that all diagonal terms (the ones in which the exponential factors cancel) are proportional to $\STr_{ab}\,\tau_a^{\alpha_2} {\bf 1}_b=0$. For the remaining terms, we want to perform the integral over $v$ by closing the integration contours in the upper or lower half-plane. The choice of half-plane is dictated by the behavior of the factors $e^{\pm i \tilde{p}_b(v) r_\ell}$, with $\tilde{p}_b(v)\sim 2v$ at large $v$. Due to the presence of the decoupling poles, the integrals are only well defined if the $u$ and $v$ contours do not intersect. Throughout this Supplemental Material, we choose to set $\mathrm{Im}\,u = +\epsilon$ for some small positive $\epsilon$ and $\mathrm{Im}\,v = 0$, as well as $\mathrm{Im}\,w = -\epsilon$ for magnons on the third mirror edge, see below. We stress that the final results do not depend on the ordering we choose for the contours, provided we keep the same for all the computations. When computing the integrals over $v$ by residues the main contribution will be from the decoupling pole at $(v,b)=(u,a)$. These contributions are what we call \textit{contact terms}. The integrand might contain other poles in the complex plane at positions $v=v_*$ but their contribution will be weighted by $e^{ -|\textrm{Im}\, \tilde{p}_b(v_*)| r_\ell}$, and will be suppressed when $r_\ell\to \infty$.

It becomes clear then that the only surviving contact term will be
\begin{align}
\label{onewrapping}
    \mathcal{C}_{(1,1,0)} &= \lim_{r_\ell\to \infty}\,\sum_{a,b=1}^\infty \int\frac{\dd u\, \dd v}{(2\pi)^2} \mu_a(u)\,\mu_b(v)\,e^{-\ell \tE_a(u) +\ii(\tilde{p}_b(v) - \tilde{p}_a(u))r_\ell}\,\frac{\STr_{ab}\,\tau^{\alpha_2}_a \,(\mathcal{S}_{ab}(u,v))^2}{h^2_{ba}(v,u)}\no \\
    &=\sum_{a=1}^\infty\int\frac{\dd u}{2\pi}e^{-\ell \tE_a(u) }\,\left(- \ii \partial_v\,\STr_{ab}\,\tau^{\alpha_2}_a \,(\mathcal{S}_{ab}(u,v))^2\right)\Big|_{v\to u}\, .
\end{align}
For the last equality, we have used \eqref{hresidue} and the fact that the terms where the $v$ derivative acts outside the super-trace are zero, since $\STr_{ab}\,\tau^{\alpha_2}_a \,(\mathcal{S}_{ab}(u,u))^2 = \STr_{ab}\,\tau_a^{\alpha_2} {\bf 1}_b=0$. This explains in particular why there are no volume-dependent corrections in the wrapping terms, which would come from the derivative acting on the exponential factor depending on $r$'s.

Before moving on to cases with more than one wrapping magnon, let us discuss the other one-wrapping magnon contributions. It is clear that $\mathcal{C}_{(0,1,1)}$ is identical to $\mathcal{C}_{(1,1,0)}$ up to the replacements $\ell \to k$ and $\alpha_2 \to \alpha_1$. For $\mathcal{C}_{(1,0,1)}$, the situation is slightly different because we need to introduce twists for both magnons: we have to compute $\STr_{ac} [ \tau^{\alpha_2}_a \, \CG(u,\emptyset,w)\, \tau^{\alpha_1}_c\, \bar\CG(u,\emptyset,w)]$. This means that the diagonal terms, i.e. those without any dependence on the large $r_p$, do not vanish. The super-trace is either directly $\STr_{ac}\,[\tau^{\alpha_2}_a\, \tau^{\alpha_1}_c] = T_a^{(\alpha_2)} T_c^{(\alpha_1)}$ or $H^{-1}_{ac}\STr_{ac}\,[\tau^{\alpha_2}_a\,\mathcal{S}_{ca}\, \tau^{\alpha_1}_c\,\mathcal{S}_{ac}] = T_a^{(\alpha_2)} T_c^{(\alpha_1)}$, using the crossed unitarity property \eqref{inverse partial transpose}. The full result is thus
\begin{equation}
    \mathcal{C}_{(1,0,1)} = \left(2\sum_{a=1}^\infty\int \frac{\dd u}{2\pi}\,e^{-\ell \tE_a(u)} \mathbb{B}^{(\alpha_2)}_1\right) \left(2\sum_{c=1}^\infty\int \frac{\dd w}{2\pi}\,e^{-k \tE_c(w)} \mathbb{B}^{(\alpha_1)}_1\right) + \sum_{a=1}^{\infty} \int \frac{\dd u}{2\pi}e^{-p \widetilde{E}_{a}(u)} \mathbb{W}^{(\alpha_1+\alpha_2)}_1\, ,
\end{equation}
where the first contribution comes from what we called the diagonal terms. Recalling the factorised form \eqref{Cfact} of the full result, we see that this contribution is a cross term coming from $\left( \! B^{(\alpha_1)}_{k} B^{(\alpha_2)}_{\ell} \!\right)^2$.

\medskip

\begin{figure}[t]
    \includegraphics[scale=0.5]{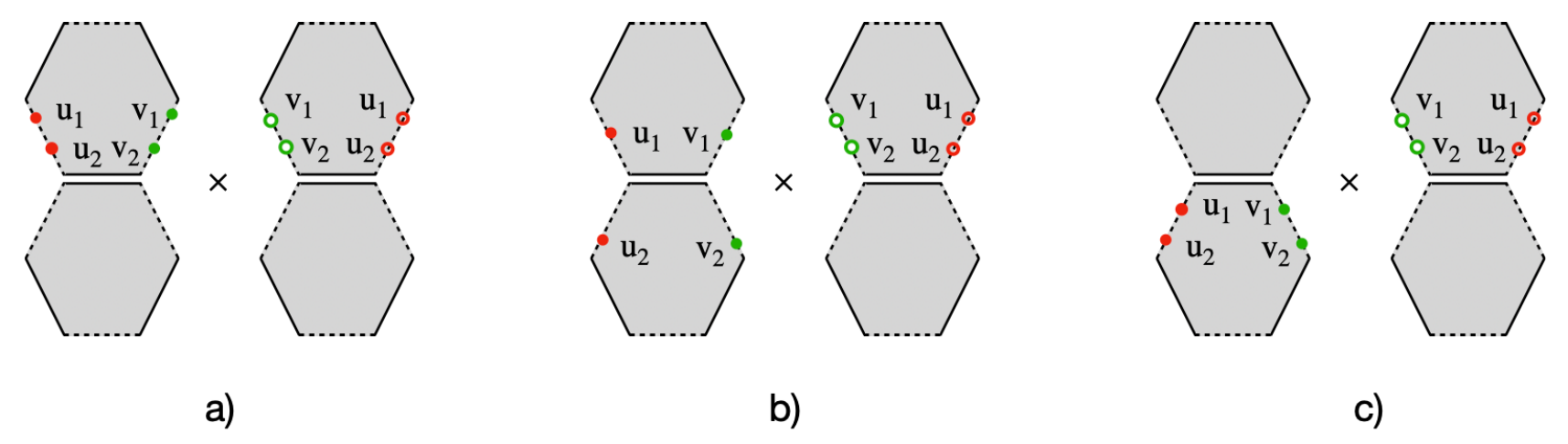}
    \caption{Different distributions of magnons that could potentially contribute to $\mathcal{C}_{(2,2,0)}$. Only the configuration in c) corresponds to a non-vanishing contribution.}
    \label{fig:WrappingTerms}
\end{figure}

{\it Multi-wrapping contributions.}
It is clear from the discussion above that the configurations with magnons on two out of the three bridges contain many terms, and we claim that they resum to \eqref{CCW}. However, in order to identify the wrapping contributions, it is enough to consider configurations with the same number $n$ of magnons on two bridges, since wrappings originate in the two-magnon contact terms. Therefore, the support of these wrapping contributions is given by $n$-fold integrals obtained by collecting the mutual poles in the same way we did above for the one-wrapping contribution. There is a unique configuration for the distribution of magnons on the four hexagons for which this is possible, illustrated in Figure \ref{fig:WrappingTerms}c) for $n=2$. We give details of the computation below. For $\mathcal{C}_{(n,n,0)}$ only the configuration with $2n$ magnons on the lower left and $2n$ magnons on the upper right hexagons survive, giving
\begin{equation}
    \! \!\mathcal{C}_{(n,n,0)}\! = \!\!\lim_{r_\ell\rightarrow \infty} \!\frac{1}{(n!)^2} \! \int \dd\u\, \dd\v \, \mu(\u)\,\mu(\v)\,e^{-\ell \widetilde{E}(\u)+\ii (\tilde{p}(\v)-\tilde{p}(\u))r_\ell} \, \frac{H(\mathbf{u}) H(\mathbf{v})}{h_{{\bf{b}}{\bf{a}}}^2(\mathbf{v},\mathbf{u})} \,\operatorname{STr} \prod_{k=1}^n \tau^{\alpha_2}_{a_k}\! \left(\prod_{i,j=1}^n  \mathcal{S}_{a_i,b_j}(u_i,v_j)\right)^{2}\!\! + \dots\, ,
\end{equation}
where we have used the Yang--Baxter equation and unitarity to simplify the matrix part, and the dots stand for all the other contributions, which we expect to vanish. Using the property \eqref{hresidue} we can perform the integration in the $v_k$'s by closing the integration contours in the upper half-plane and picking the residues at some decoupling poles. We point out that two rapidities $v_i$ and $v_k$ cannot decouple to the same rapidity $u_j$. This happens because of the factor $H(\mathbf{v})$ in the numerator and because the trace vanishes when two $(b_i,v_i)$ and $(b_k,v_k)$ are equal to the same $(a_j,u_j)$. There are thus $n!$ ways of identifying the $u_j = v_k$ and all of them are equivalent because the integrand is completely symmetric in the $u_j$'s and, separately, in the $v_k$'s. The final result is
\begin{equation}\label{Cnn0 dots}
    \mathcal{C}_{(n,n,0)} = \frac{1}{n!} \prod_{k=1}^n \left(\sum_{a_k=1}^{\infty} \int \frac{\dd u_k}{2\pi}e^{-\ell \widetilde{E}_{a_k}(u_k)}\right) \mathbb{W}^{(\alpha_2)}_n + \dots\, ,
\end{equation}
where
\begin{equation}
    \mathbb{W}^{(\alpha_2)}_n = \operatorname{STr} \left[\prod_{k=1}^n \! \tau^{\alpha_2}_{a_k}(-\ii\, \p_{v_k})\! \left(\prod_{i=1}^n\! \prod_{j=1}^n \mathcal{S}_{a_i,b_j}(u_i,v_j)\!\right)^2 \right] \Bigg|_{\,{\bf v}\to {\bf u}}\, .
\end{equation}

The rest of this subsection is devoted to discussing the terms contained in the dots in \eqref{Cnn0 dots}. We will not show that they all vanish, but there are some simple arguments that explain why many of them do. We first note that we need to bring down an even number of magnons in order to have a chance of cancelling the exponential factors $e^{\pm\ii\tilde{p} r}$ by picking the residues of decoupling poles. Then, we note that all the diagonal terms, i.e. those that have the same magnon configurations on the left and right pairs of pants, vanish because the S-matrices cancel and the supertrace over the $v$ magnons yields zero. Further, we remark that if the integrand contains a factor $e^{+\ii\tilde{p}_a(u)r}$ (respectively $e^{-\ii\tilde{p}_c(w)r}$) then the integral over $u$ (respectively $w$) would be suppressed because our ordering of the contours $\mathrm{Im}(u) > \mathrm{Im}(v) >\mathrm{Im}(w)$ means that there could not be any decoupling pole whose residue to pick.

Let us focus a bit more on the example of $\mathcal{C}_{(2,2,0)}$. One of the diagonal terms is drawn in Figure \ref{fig:WrappingTerms}a). In this configuration where all the magnons are in the upper hexagons, the contribution from the form factors and twists is
\begin{equation}\label{toptop}
    \frac{H_{u_1 u_2} H_{v_1 v_2}}{\prod_{j,k=1}^2 H_{u_j v_k}} \STr\,[\tau^{\alpha_2}_{u_1}\,\tau^{\alpha_2}_{u_2}\,(\CS_{u_1 u_2}\,\CS_{v_1 u_2}\,\CS_{v_2 u_2}\,\CS_{v_1 u_1}\,\CS_{v_2 u_1}\,\CS_{v_2 v_1}) \, (\CS_{v_1 v_2}\,\CS_{u_1 v_2}\,\CS_{u_1 v_1}\,\CS_{u_2 v_2}\,\CS_{u_2 v_1}\,\CS_{u_2 u_1})]\, .
\end{equation}
Using unitarity of the S-matrix, this simply reduces to
\begin{equation}
    \frac{H_{u_1 u_2} H_{v_1 v_2}}{\prod_{j,k=1}^2 H_{u_j v_k}} \STr\,[\tau^{\alpha_2}_{u_1}\,\tau^{\alpha_2}_{u_2}]=0
\end{equation}
because of the super-trace over the spaces associated to $v_1$ and $v_2$, as claimed above. It is then easy to see that the general arguments we gave above exclude all configurations associated with $\mathcal{C}_{(2,2,0)}$ except those drawn in Figure \ref{fig:WrappingTerms}b) (up to permutations $u_1\leftrightarrow u_2$ and or $v_1\leftrightarrow v_2$) and Figure \ref{fig:WrappingTerms}c). The former corresponds to
\begin{equation}\label{bridge wrap}
    e^{\ii(\tilde{p}_{b_2}(v_2) - \tilde{p}_{a_2}(u_2))} \frac{h_{u_2 u_1} h_{v_1 v_2}}{H_{u_1 v_1} h^2_{v_2 u_2} h_{v_1 u_2} h_{v_2 u_1}}\STr\,[\tau^{\alpha_2}_{u_1}\,\tau^{\alpha_2}_{u_2}\,(\CS_{v_1 u_1} \, \CS_{u_2 v_2} )\,(\CS_{v_1 v_2}\,\CS_{u_1 v_2}\,\CS_{u_1 v_1}\,\CS_{u_2 v_2}\,\CS_{u_2 v_1}\,\CS_{u_2 u_1})]\, .
\end{equation} 
Given the form of the exponential factor and the relative ordering of the contours, we need to perform the integral in $v_2$ by picking the residue of the double pole at $v_2 = u_2$. As before, the terms for which the derivative acts on the scalar part vanish. Indeed, using $\mathcal{S}_{ab}(u,u) = \mathcal{P}_{ab}$ the expression above becomes proportional to
\begin{align}
    \STr\,[\tau^{\alpha_2}_{u_1}\,\tau^{\alpha_2}_{u_2}\,\CS_{v_1 u_1} \,\CS_{v_1 u_2}\,\CS_{u_1 u_2}\,\CS_{u_1 v_1}\,\CS_{u_2 v_1}\,\CS_{u_2 u_1}] = \STr\,[\tau^{\alpha_2}_{u_1}\,\tau^{\alpha_2}_{u_2}] = 0\, ,
\end{align}
where we used the Yang--Baxter relation and unitarity. A similar argument holds when the derivative acts on $\mathcal{S}_{u_1 v_2}$ or on $\mathcal{S}_{v_1 v_2}$. And if it acts on the first $\mathcal{S}_{u_2 v_2}$, we get
\begin{equation}
    \frac{1}{H_{u_1 v_1}}\STr\,[\tau^{\alpha_2}_{u_1}\,\tau^{\alpha_2}_{u_2}\,\CS_{v_1 u_1} \left(\partial_{2}\CS_{u_2 v_2}\right) \mathcal{P}_{u_2 v_2}\,\CS_{u_1 v_1}]  = \frac{1}{H_{u_1 v_1}}\STr\,[\tau^{\alpha_2}_{u_1}\,\tau^{\alpha_2}_{u_2} \, \left(\partial_{2}\CS_{u_2 v_2}\right) \mathcal{P}_{u_2 v_2}] = 0
\end{equation}
because of the super-trace over the space associated to $v_1$. If the derivative acts on the other instance of $\mathcal{S}_{u_2 v_2}$, the same result holds. We have thus verified that only \ref{fig:WrappingTerms}c) gives a non-zero contribution to $\mathcal{C}_{(2,2,0)}$, so that
\begin{equation}
    \mathcal{C}_{(2,2,0)} = \frac{1}{2} \prod_{k=1}^2 \left(\sum_{a_k=1}^{\infty} \int \frac{\dd u_k}{2\pi}e^{-\ell \widetilde{E}_{a_k}(u_k)}\right) \mathbb{W}^{(\alpha_2)}_2 \, .
\end{equation}

Finally, let us discuss the case of $\mathcal{C}_{(2,0,2)}$ where the factorized form of the full result \eqref{3ptBPS} can be probed. In this case, the wrapping term is of course still present (with twist $\alpha_3 = \alpha_1+\alpha_2$ and bridge length $p = k+ \ell$) but the other terms do not all vanish anymore. For instance, the analogue of \eqref{toptop} now reads
\begin{equation}
    \frac{H_{u_1 u_2} H_{w_1 w_2}}{\prod_{j,k=1}^2 H_{u_j w_k}} \STr\,[\tau^{\alpha_2}_{u_1}\,\tau^{\alpha_2}_{u_2}\,(\CS_{u_1 u_2}\,\CS_{w_1 u_2}\,\CS_{w_2 u_2}\,\CS_{w_1 u_1}\,\CS_{w_2 u_1}\,\CS_{w_2 w_1}) \,\tau^{\alpha_1}_{w_1}\,\tau^{\alpha_1}_{w_2}\, (\CS_{w_1 w_2}\,\CS_{u_1 w_2}\,\CS_{u_1 w_1}\,\CS_{u_2 w_2}\,\CS_{u_2 w_1}\,\CS_{u_2 u_1})]\, .
\end{equation}
Using unitarity of the S-matrix as well as the property \eqref{partial trace M}, this simply reduces to
\begin{equation}
    H_{u_1 u_2} H_{w_1 w_2} \STr\,[\tau^{\alpha_2}_{u_1}\,\tau^{\alpha_2}_{u_2}\,\tau^{\alpha_1}_{w_1}\,\tau^{\alpha_1}_{w_2}] = H_{u_1 u_2} H_{w_1 w_2}  T_{a_1}^{(\alpha_2)} T_{a_2}^{(\alpha_2)} T_{c_1}^{(\alpha_1)} T_{c_2}^{(\alpha_1)} = \frac{\mathbb{B}^{(\alpha_2)}_2(\u)\,\mathbb{B}^{(\alpha_1)}_2(\w)}{\mu(\u)\, \mu(\w)}\, .
\end{equation}
Similarly, the analogue of \eqref{bridge wrap} reads
\begin{equation}
    e^{\ii(\tilde{p}_{c_2}(w_2) - \tilde{p}_{a_2}(u_2))} \frac{h_{u_2 u_1} h_{w_1 w_2}}{H_{u_1 w_1} h^2_{w_2 u_2} h_{w_1 u_2} h_{w_2 u_1}} \STr\,[\tau^{\alpha_2}_{u_1}\,\tau^{\alpha_2}_{u_2}\,(\CS_{w_1 u_1} \, \CS_{u_2 w_2})\, \tau^{\alpha_1}_{w_1}\,\tau^{\alpha_1}_{w_2}\, (\CS_{w_1 w_2}\,\CS_{u_1 w_2}\,\CS_{u_1 w_1}\,\CS_{u_2 w_2}\,\CS_{u_2 w_1}\,\CS_{u_2 u_1})]\, .
\end{equation}
We need to perform the integral in $w_2$ by picking the residue of the double pole at $w_2 = u_2$. As usual, the term for which the derivative acts on the scalar part can be shown to vanish. However, when the derivative acts on the S-matrices we get
\begin{multline}
    \frac{1}{H_{u_1w_1}} (-\ii\,\p_{w_2}) \STr\,[\tau^{\alpha_2}_{u_1}\,\tau^{\alpha_2}_{u_2}\,(\CS_{w_1 u_1} \, \CS_{u_2 w_2})\, \tau^{\alpha_1}_{w_1}\,\tau^{\alpha_1}_{w_2}\, (\CS_{w_1 w_2}\,\CS_{u_1 w_2}\,\CS_{u_1 w_1}\,\CS_{u_2 w_2}\,\CS_{u_2 w_1}\,\CS_{u_2 u_1})]\big|_{w_2\to u_2}\\
    = T_{a_1}^{(\alpha_2)} T_{c_1}^{(\alpha_1)} \,\mathbb{W}^{(\alpha_1+\alpha_2)}_1 (u_2) = \frac{\mathbb{B}^{(\alpha_2)}_1(u_1) \,\mathbb{B}^{(\alpha_1)}_1(w_1)}{\mu(u_1) \,\mu(w_1)} \,\mathbb{W}^{(\alpha_1+\alpha_2)}_1 (u_2)\, ,
\end{multline}
where the first equality relies on the partial trace formula \eqref{partial traces derivatives}.

\subsubsection*{Bridge-like contribution} 
     
\begin{figure}[t]
    \includegraphics[scale=0.38]{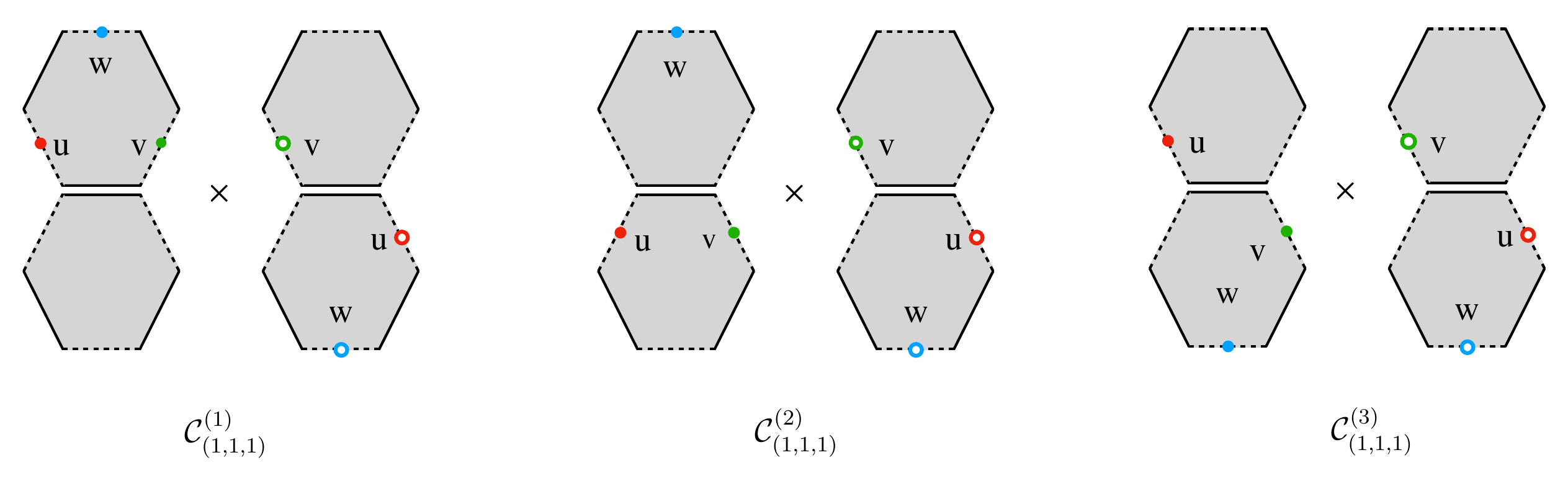}
    \caption{Three three-magnon configurations giving rise to the bridge-like contribution $\mathcal{C}_{(1,1,1)}^{(j)}$, $j=1,2,3$. The configurations rotated by $180^\circ$ contribute the same amount. The pairs of  magnons $(u,w)$, $(u,v)$ and $(w,v)$  are transported through the legs of lengths $r_{l+k}$, $r_\ell$ and $r_k$ respectively.}
\label{fig:BridgeLike}
\end{figure}

{\it One-magnon bridge-like contribution.}
There is another type of contact terms coming from identifying magnons in the three mirror bridges of a hexagon, $(u_i,a_i) = (v_j,b_j) = (w_k,c_k)$. We consider first the simplest case $\mathcal{C}_{(1,1,1)}$. From the previous discussions, we understand that the configurations that survive at large $r$'s are those where pairs of magnons on opposite edges are transported through the same leg. Some of them are depicted in Figure \ref{fig:BridgeLike}. The first configuration evaluates to
\begin{multline}
    \mathcal{C}^{(1)}_{(1,1,1)} = \lim_{r_{p}\rightarrow \infty} \sum_{a, b, c = 1}^{\infty} \int \frac{\dd u\, \dd v\, \dd w}{(2\pi)^3}\,\mu_a(u)\, \mu_b(v)\, \mu_c(w) \, e^{-\ell \widetilde{E}_a(u) -k \widetilde{E}_c(w)+ \ii(\tilde{p}_c(w)-\tilde{p}_a(u))r_{p}}\,h_{ac}(u^{4\gamma},w)\,h_{ba}(v^{4\gamma},u)\\
    \times h_{ca}(w^{2\gamma},u)\,h_{bc}(v^{4\gamma},w^{2\gamma})\, \STr_{abc}\left[\tau_a^{\alpha_2}\, \mathcal{S}_{ca}(w^{2\gamma},u)\,\mathcal{S}_{ba}(v^{4\gamma},u)\,\mathcal{S}_{bc}(v^{4\gamma},w^{2\gamma})\,\tau_c^{\alpha_1}\,\mathcal{S}_{ac}(u^{4\gamma},w)\right]\, .
\end{multline}

Using the properties \eqref{huvproperties} and \eqref{gammasS} of the dynamical factor and the S-matrix, we can bring the previous expression to the form
\begin{multline}
    \mathcal{C}^{(1)}_{(1,1,1)} = \lim_{r_{p}\rightarrow \infty} \sum_{a, b, c = 1}^{\infty} \int \frac{\dd u\, \dd v\, \dd w}{(2\pi)^3}\,\mu_a(u)\, \mu_b(v)\, \mu_c(w) \, e^{-\ell \widetilde{E}_a(u) -k \widetilde{E}_c(w) + \ii(\tilde{p}_c(w)-\tilde{p}_a(u))r_{p}}\\
    \times\frac{h_{ca}(w^{2\gamma},u)}{h_{ab}(u,v) \,h_{cb}(w^{2\gamma},v)\, h_{c a}(w,u)} \, \STr_{abc}\left[ \tau^{\alpha_2}_a\, \mathcal{S}_{ca}(w^{2\gamma},u)\,\mathcal{S}_{ba}(v,u)\,\mathcal{S}_{bc}(v,w^{2\gamma})\,\tau_c^{\alpha_1} \, \kappa_c\,\mathcal{S}_{ac}(u,w)\,\kappa_c \right]\, .
\end{multline}
Now we perform the integral over $u$ by deforming the contour in the lower half-plane and picking the residue at $(a,u) = (b,v)$. Using $\mathcal{S}_{aa}(u,u) =  \mathcal{P}^{g}$  and unitarity, the super-trace simplifies and the result is given by
\begin{align}
    \mathcal{C}^{(1)}_{(1,1,1)} = -\lim_{r_{p}\rightarrow \infty} \,\sum_{b,c=1}^{\infty} \int \frac{\dd v\, \dd w}{(2\pi)^2} \mu_b(v) \mu_c(w) e^{-\ell \widetilde{E}_b(v) -k \widetilde{E}_c(w) + \ii(\tilde{p}_c(w)-\tilde{p}_b(v))r_{p}} \frac{1}{h_{cb}(w,v)} \STr_{bc} \left[\tau_b^{\alpha_2} \mathcal{S}_{bc}(u,w) \tau_c^{\alpha_1}\right]\, .
\end{align}
Then, performing the integral over $v$ by closing the contour in the lower half-plane and picking the residue of the pole in $(b,v) = (c,w)$, we arrive at
\begin{align}
    \mathcal{C}^{(1)}_{(1,1,1)} = -\sum_{c=1}^{\infty} T^{(\alpha_3)}_c \int \frac{\dd w}{2\pi}e^{-p \widetilde{E}_c(w)} \mu_c(w)\, .
\end{align}
When we computed the integral over $u$, we should have also picked the residue of the apparent double pole at $(a,u) = (c,w)$. However, one can use \eqref{partial traces} to show that it vanishes.

There are other configurations that contribute. Two of them are obtained from the one we have already considered by bringing down either the pair $(u,v)$ or the pair $(v,w)$ on the left pair of pants, see Figure \ref{fig:BridgeLike}. The first one produces
\begin{multline}
    \mathcal{C}^{(2)}_{(1,1,1)} = \lim_{r_\ell,r_{p}\rightarrow \infty} \sum_{a, b, c = 1}^{\infty} \int \frac{\dd u\, \dd v\, \dd w}{(2\pi)^3}\,\mu_a(u)\, \mu_b(v)\, \mu_c(w) \, e^{-\ell \widetilde{E}_a(u) -k \widetilde{E}_c(w) + \ii(\tilde{p}_c(w)-\tilde{p}_a(u))r_{p} + \ii(\tilde{p}_b(v)-\tilde{p}_a(u))r_{\ell}} \\
    \times \frac{\STr_{abc}\, \tau^{\alpha_2}_a \, \mathcal{S}_{ab}(u,v) \, \tau^{\alpha_1}_c \, \mathcal{S}_{ac}(u,w)}{h_{ab}(v,u)\, h_{ca}(w,u)} = \sum_{a=1}^{\infty} T^{(\alpha_3)}_a \int \frac{\dd u}{2\pi}e^{-p \widetilde{E}_a(u)} \mu_a(u)\, ,
\end{multline}
where we computed the integrals over $v$ and $w$ by closing the contours in the upper half-plane and picking the residues of the simple decoupling poles at $(b,v) = (a,u)$ and $(c,w) = (a,u)$. The second configuration yields the same result, 
\begin{equation}
    \mathcal{C}^{(3)}_{(1,1,1)} = \mathcal{C}^{(2)}_{(1,1,1)} = -\mathcal{C}^{(1)}_{(1,1,1)}\,.
\end{equation}

The last three non-vanishing configurations can be obtained graphically by rotating the previous three by $180$°, hence they are equal to those we have previously computed. Summing up the six non-zero configurations we obtain indeed the first non-trivial term in the expansion of $\left(\!B^{(\alpha_3)}_{p}\right)^2$, namely
\begin{align}
    \mathcal{C}_{(1,1,1)} = 2\left(\mathcal{C}^{(1)}_{(1,1,1)}+\mathcal{C}^{(2)}_{(1,1,1)}+\mathcal{C}^{(3)}_{(1,1,1)}\right)=2\,\mathcal{C}^{(2)}_{(1,1,1)}=2\sum_{a=1}^\infty\int \frac{\dd u}{2\pi}\,e^{-p \tE_a(u)}\, \mathbb{B}^{(\alpha_3)}_1 \, .
\end{align}
By inspection, these are the only non-vanishing terms where the three rapidities can be identified. Other terms with non-coinciding rapidities can be studied in the same way as in the previous sections and reproduce parts of either $\left(\!B^{(\alpha_1)}_{k} W^{(\alpha_2)}_{\ell}\right)^2$ or $\left(\!B^{(\alpha_2)}_{\ell} W^{(\alpha_1)}_{k}\right)^2$.
    
{\it Multi-magnon bridge-like contributions.}
We remark that although it is difficult to explicitly compute the higher bridge-like contributions, they will come from configurations which are superpositions of copies of the diagrams of Figure \ref{fig:BridgeLike} and their rotated counterparts. As an example, let us discuss briefly the case of $\mathcal{C}_{(2,2,2)}$. Consider first the diagram formed by superposing two copies of $\mathcal{C}^{(2)}_{(1,1,1)}$ in Figure \ref{fig:BridgeLike}. The hexagons together with the combinatorial factors in \eqref{Cpants} give, after using the properties of the S-matrices,
\begin{align}
   \frac{H_{u_1 u_2} H_{v_1 v_2} H_{w_1 w_2}}{(2!)^3 \prod^2_{i,j=1} h_{v_i u_j} h_{w_i u_j}}\STr\left[\tau^{\alpha_2}_{u_1} \tau^{\alpha_2}_{u_2}\,  \mathcal{S}_{u_2 v_1} \mathcal{S}_{u_2 v_2} \mathcal{S}_{u_1 v_1} \mathcal{S}_{u_1 v_2}   \,\tau^{\alpha_1}_{w_1} \tau^{\alpha_1}_{w_2}\, \mathcal{S}_{u_2 w_1} \mathcal{S}_{u_2 w_2} \mathcal{S}_{u_1 w_1} \mathcal{S}_{u_1 w_2} \right]\, .
\end{align}
There are $4$ ways of identifying the three sets of rapidities, all of which yield the same result. Choosing for example $(a_j,u_j) = (b_j,v_j) = (c_j,w_j)$, we get:
\begin{align}\label{two bridge-like example}
    \frac{H_{u_1 u_2}}{2} \STr\left[\tau^{\alpha_2}_{u_1} \tau^{\alpha_2}_{u_2}\,  \mathcal{S}_{u_2 v_1} \mathcal{P}_{u_2 v_2} \mathcal{P}_{u_1 v_1} \mathcal{S}_{u_1 v_2} \,  \tau^{\alpha_1}_{w_1} \tau^{\alpha_1}_{w_2}\, \mathcal{S}_{u_2 w_1} \mathcal{P}_{u_2 w_2} \mathcal{P}_{u_1 w_1} \mathcal{S}_{u_1 w_2} \right] = \frac{H_{u_1 u_2}}{2}\, T^{(\alpha_3)}_{a_1} T^{(\alpha_3)}_{a_2} = \frac{\mathbb{B}^{(\alpha_3)}_2(\u)}{2\,\mu(\u)}\, .
\end{align}
In total, there are $9$ ways of superposing two configurations of $\mathcal{C}_{\text{up}}= \{\mathcal{C}^{(1)}_{(1,1,1)},\mathcal{C}^{(2)}_{(1,1,1)},\mathcal{C}^{(3)}_{(1,1,1)}\}$, their contributions sum to \eqref{two bridge-like example}. We can also superpose the diagrams we obtain rotating $\mathcal{C}_{\text{up}}$ by $180^\circ$, which we call $\mathcal{C}_{\text{up}}^{\text{rotated}}$, with themselves to get an identical contribution. Finally, we can superpose $\mathcal{C}_{\text{up}}$ and $\mathcal{C}_{\text{up}}^{\text{rotated}}$, which will produce the square of the one-magnon part of the bridge-like contribution $\det \left(1-s_{\alpha_3} K_{p}\right)$; this is expected since we need to reproduce $(\det \left(1-s_{\alpha_3} K_{p}\right))^2$. For the full structure constant to be factorized, $\mathcal{C}_{(2,2,2)}$ should also contain many other terms. We do not describe all of them here but we will consider a slightly simpler example in detail in the next subsection.

\subsubsection*{More on factorization}

\begin{figure}[t]
    \includegraphics[scale=0.38]{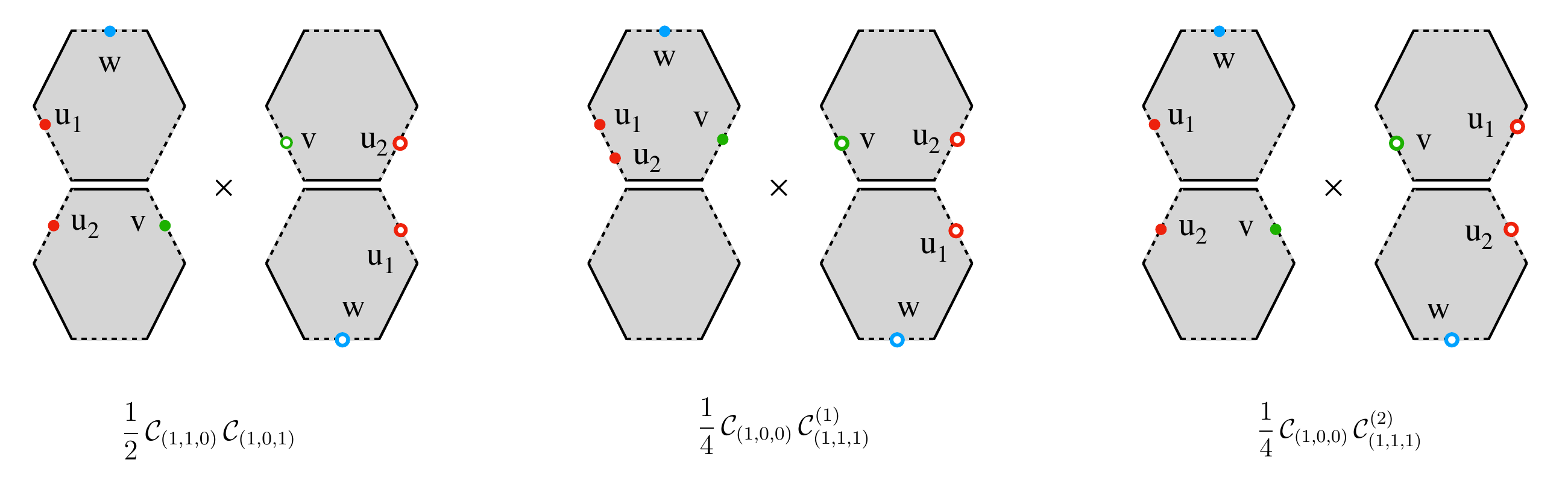}
    \caption{Three different configurations contributing to $\mathcal{C}_{(2,1,1)} = \mathcal{C}_{(1,1,0)}\,\mathcal{C}_{(1,0,1)} + \mathcal{C}_{(1,0,0)}\,\mathcal{C}_{(1,1,1)}$. The first one factorizes into two wrapping magnons, the other two factorize into different contributions to the bridge $\times$ bridge-like contribution. A factor of two arises from the exchange of $u_1$ and $u_2$. In the two rightmost configurations, another factor of two arises from putting the bridge magnon in the lower hexagons. }
    \label{fig:FactorizedMagnons}
\end{figure}

An important feature of the structure constants considered in this work is their factorization in different pieces: bridge, wrapping, and bridge-like. Although we do not have an argument valid for any number of magnons, we have some understanding of how it occurs for low number of magnons. We already saw a few examples of cross terms between bridge and wrapping contributions in the previous two subsections. We examine here one more example, $\mathcal{C}_{(2,1,1)}$, which contains a cross term between a bridge and a bridge-like magnon. As should be clear, these cross terms arise from diagrams that are superpositions of one bridge diagram and a bridge-like one; see Figure \ref{fig:FactorizedMagnons} middle and right. Let us begin with the former. The relevant part of the integrand is
\begin{align}
    e^{\ii(\tilde{p}_c(w)-\tilde{p}_{a_1}(u_1))r_{p}} \frac{h_{u_1 u_2}\, h^2_{w^{2\gamma} u_1}\, h_{w^{2\gamma} u_2}\, h_{v^{2\gamma} w}}{2\, h_{u_1 v}\, h_{u_2 v}\, h_{v u_2}} \STr\left[\tau^{\alpha_2}_{u_1} \tau^{\alpha_2}_{u_2}\,\mathcal{S}_{u_1 u_2} \mathcal{S}_{w^{2\gamma} u_2} \mathcal{S}_{v u_2}\, \mathcal{S}_{w^{2\gamma} u_1} \mathcal{S}_{v u_1} \mathcal{S}_{v w^{2\gamma}}\, \tau^{\alpha_1}_{w}\,\mathcal{S}_{w^{2\gamma} u_1} \mathcal{S}_{u_2 v}\right]\, .
\end{align}
We first perform the integral over $u_1$ by closing the contour in the lower half-plane. Only the residue of the pole at $u_1 = v$ contributes, and it gives, after using the properties \eqref{inverse} and \eqref{partial trace perm} of the S-matrices, 
\begin{multline}\label{factorisation detail}
    -\ii\, e^{\ii(\tilde{p}_c(w)-\tilde{p}_{b}(v))r_{p}} \frac{h_{w^{2\gamma} u_2}\, h_{w^{2\gamma} v}\, h_{v u_2}}{2\, \mu_{v}} \STr\left[\tau^{\alpha_2}_{v} \tau^{\alpha_2}_{u_2}\,\mathcal{S}_{v u_2} \mathcal{S}_{w^{2\gamma} u_2}\, \tau^{\alpha_1}_{w}\,\mathcal{S}_{w^{2\gamma} v}\right]\\
    = -\ii\, e^{\ii(\tilde{p}_c(w)-\tilde{p}_{b}(v))r_{p}} \frac{h_{v u_2}}{2\,h_{w u_2}\, h_{w v}\, \mu_{v}} \STr\left[\tau^{\alpha_2}_{v} \tau^{\alpha_2}_{u_2}\,\tau^{\alpha_1}_{w}\, \mathcal{S}_{v u_2}\, \mathcal{S}_{u_2 w}\, \mathcal{S}_{v w}\right]\, ,
\end{multline}
where we used the crossing property \eqref{CrossingS} to rewrite the super-trace.  Integrating now over $w$ means taking the residue of the pole at $w = v$ to get
\begin{equation}
    -\frac{1}{2\, \mu^2_{v}} \STr\left[\tau^{\alpha_3}_{u_1} \tau^{\alpha_2}_{u_2}  \right] = -\frac{1}{2\, \mu^2_{v}} T^{(\alpha_2)}_{a_2} T^{(\alpha_3)}_{b}\, ,
\end{equation}
which agrees with our expectation since it is clearly a piece of $\left(\!B^{(\alpha_2)}_\ell B^{(\alpha_3)}_p\right)^2$.

The relevant part of the integrand associated with Figure \ref{fig:FactorizedMagnons} right is
\begin{equation}
    e^{\ii(\tilde{p}_b(v)-\tilde{p}_{a_2}(u_2))r_{\ell}+\ii(\tilde{p}_c(w)-\tilde{p}_{a_2}(u_2))r_{p}} \frac{h_{w^{2\gamma} u_1}\, h_{w^{2\gamma} u_2} \, h_{u_2 u_1}}{2\, h_{v u_1}\, h_{v u_2}\, h_{u_1 u_2}}\STr\left[ \tau^{\alpha_2}_{u_1}\,\tau^{\alpha_2}_{u_2}\,\mathcal{S}_{w^{2\gamma} u_1}\, \mathcal{S}_{u_2 v}\, \tau_w^{\alpha_1}\, \mathcal{S}_{u_2 u_1}\, \mathcal{S}_{u_1 v}\, \mathcal{S}_{w^{2\gamma} u_2}\, \mathcal{S}_{u_2 u_1}\right]\, ,
\end{equation}
where the two $\mathcal{S}_{u_1 u_2}$ and part of the dynamical factor come from the transition factor \eqref{transitionf} because we crossed magnon $u_1$ while bringing $u_2$ down the right pair of pants. Integrating over $v$ means taking the residue at $v=u_2$ to get 
\begin{equation}
    -\ii\,e^{\ii(\tilde{p}_c(w)-\tilde{p}_{a_2}(u_2))r_{p}} \frac{h_{w^{2\gamma} u_1}\, h_{w^{2\gamma} u_2}\, h_{u_2 u_1}}{2\, \mu_{u_2}}\STr\left[ \tau^{\alpha_2}_{u_1}\,\tau^{\alpha_2}_{u_2}\,\mathcal{S}_{w^{2\gamma} u_1}\, \tau_w^{\alpha_1}\,\mathcal{S}_{w^{2\gamma} u_2}\, \mathcal{S}_{u_2 u_1}\right]\, .
\end{equation}
This is the same as \eqref{factorisation detail} up to the replacement $(v,u_2)\to(u_2,u_1)$, hence the residue at $w=u_2$ is also given by $-T^{(\alpha_2)}_{a_2} T^{(\alpha_3)}_{b}/2\mu^2_{u_2}$. However, the two contributions will come with opposite relative sign because in the case of Figure \ref{fig:FactorizedMagnons} center we closed one contour in the upper half-plane and one in the lower half-plane, and in the case of Figure \ref{fig:FactorizedMagnons} right, we closed both contours in the upper half-plane.

Another example of factorization is Figure \ref{fig:FactorizedMagnons} left. There, the pairs of magnons $(w,u_1)$ and $(v,u_2)$ do not interact and each of them generates a different wrapping contribution, yielding a piece of $\left(\!W^{(\alpha_2)}_{\ell}W^{(\alpha_3)}_{p}\right)^2$. Considering all the possible diagrams, we finally find
\begin{multline}
    \mathcal{C}_{(2,1,1)} = \left(\sum_{a=1}^\infty\int \frac{\dd u}{2\pi}\,e^{-\ell \tE_a(u)} \mathbb{W}^{(\alpha_2)}_1\right) \left(\sum_{c=1}^\infty\int \frac{\dd w}{2\pi}\,e^{-p \tE_c(w)} \mathbb{W}^{(\alpha_3)}_1\right)\\
    + \left(2\sum_{a=1}^\infty\int \frac{\dd u}{2\pi}\,e^{-\ell \tE_a(u)} \mathbb{B}^{(\alpha_2)}_1\right) \left(2\sum_{c=1}^\infty\int \frac{\dd w}{2\pi}\,e^{-p \tE_c(w)} \mathbb{B}^{(\alpha_3)}_1\right)\,.
\end{multline}

\subsection*{Pfaffian and connected terms}
   
In this section we remind a few facts about the expansion of the Pfaffian in cycles similar to the one in \cite{Kostov:2019auq}. Consider the antisymmetric $2n\times 2n$ matrix $\K$ with elements
\begin{align}
    \K_{ij}^{\epsilon_i,\epsilon_j} = -\K_{ji}^{\epsilon_j,\epsilon_i} = \frac{x_i^{[\epsilon_i]}-x_j^{[\epsilon_j]}}{x_i^{[\epsilon_i]}x_j^{[\epsilon_j]}-1}\,, \quad 1\leqslant i,j \leqslant n\, , \quad \epsilon_i,\epsilon_j=\pm\, .
\end{align}
Define
\begin{align}
    \H_{ij}=\prod_{\epsilon_i,\epsilon_j=\pm} \K_{ij}^{\epsilon_i,\epsilon_j} \qquad \text{and}\qquad \H_{i}=\K_{ii}^{+-}\,.
\end{align}
Here, for simplicity, we consider just the matrices for the fundamental states, $a_i=1$, and, unlike in the main text, the indices for the matrices refer to the rapidity rather than the bound state. The Pfaffian of $\K$ can be written as 
\begin{align}
    \textrm{Pf}_n(\K)=\prod_{i=1}^n \H_i \,\prod_{1\leqslant i<j\leqslant n}\H_{ij}\, .
\end{align}
Defining a cycle of length $n$ as     
\begin{align}
    (\K_{12}\,\K_{23}\ldots \,\K_{n1})\equiv \frac{1}{2}\sum_{\epsilon_j=\pm}(\prod_j \epsilon _j )\,\K^{-\epsilon_1,\epsilon_2}_{12}\,\K^{-\epsilon_2,\epsilon_3}_{23}\ldots \,\K^{-\epsilon_n,\epsilon_1}_{n1}\,,\qquad (\K_{11})\equiv -\H_1\,,
\end{align}
and using the expression of the Pfaffian as a sum over permutations, one can express the Pfaffian as sums of products of cycles (connected terms)
\begin{align}
\label{cycleex}
    \Pf_1(\K)&=-(\K_{11})\,,\\\no
    \Pf_2(\K)&=(\K_{11})\,(\K_{22})-(\K_{12}\,\K_{21})\,,\\\no
    \Pf_3(\K)&=-(\K_{11})\,(\K_{22})\,(\K_{33})+(\K_{11})\,(\K_{23}\,\K_{32})+(\K_{22})\,(\K_{13}\,\K_{31})+(\K_{33})\,(\K_{12}\,\K_{21})\\ \no
    &-(\K_{12}\,\K_{23}\,\K_{31})-(\K_{13}\,\K_{32}\,\K_{21})\,.
\end{align}
A cycle is invariant over circular permutation of indices, {\it e.g.} $(\K_{12}\,\K_{21})=(\K_{21}\,\K_{12})$. Taking by convention the beginning of the cycle at index $1$, there will be $(n-1)!$ different cycles of length $n$ contributing to the connected part. Since all these different cycles give the same answer after integration, it is natural to define the connected part by dividing by the combinatorial factor $(n-1)!$. The relations \eqref{cycleex} can then be inverted to get the expression of the of the connected part in terms of the weights $\H_{ij}$ 
\begin{align}
\label{cycletoH}
    \textrm{CPf}_1(\K)&\equiv-(\K_{11})=\H_1\,,\\\no
    \textrm{CPf}_2(\K)&\equiv-(\K_{12}\,\K_{21})=\H_1\,\H_2\,( \H_{12}-1) \,,\\\no
    \textrm{CPf}_3(\K)&\equiv\frac{1}{2}\left[ -(\K_{12}\,\K_{23}\,\K_{31})-(\K_{13}\,\K_{32}\,\K_{21})\right]=\frac{1}{2}\H_1\,\H_2\,\H_3\,(\H_{12}\,\H_{13}\,\H_{23}-\H_{12}-\H_{13}-\H_{23}+2)\,.
\end{align}

\end{document}